\documentclass[manuscript=article,layout=twocolumn]{achemso}
\usepackage{natbib}
\usepackage{xr}
\makeatletter
\newcommand*{\addFileDependency}[1]{
  \typeout{(#1)}
  \@addtofilelist{#1}
  \IfFileExists{#1}{}{\typeout{No file #1.}}
}
\makeatother

\usepackage[version=3]{mhchem} 
\usepackage{siunitx}
\usepackage{amssymb}
\usepackage{amsmath}
\usepackage{lipsum}
\usepackage{multicol}
\usepackage[labelfont=bf]{caption} 
\usepackage{booktabs,caption}
\usepackage{threeparttable}
\usepackage{titlesec,hyperref}
\usepackage{nameref}
\usepackage{color}
\usepackage{sidecap, caption}
\usepackage{chemfig}
\usepackage{adjustbox}
\usepackage{graphicx}
\usepackage{placeins}
\usepackage{chemfig}
\usepackage{array}
\usepackage{comment}
\usepackage{tabularx,booktabs}
\DeclareSIUnit\bar{bar}

\newcommand{\beginsupplement}{%
        \setcounter{table}{0}
        \renewcommand{\thetable}{S\arabic{table}}%
        \setcounter{figure}{0}
        \renewcommand{\thefigure}{S\arabic{figure}}%
        \setcounter{equation}{0}
        \renewcommand{\theequation}{S\arabic{equation}}%
     }
\newcommand{\blackline}{\raisebox{2pt}{\protect\tikz{\protect\draw[-,black,solid,line width = 0.9pt](0,0) -- (5mm,0);}}}
\newcommand{\blueline}{\raisebox{2pt}{\protect\tikz{\protect\draw[-,blue,solid,line width = 0.9pt](0,0) -- (5mm,0);}}}

\newcommand{\redline}{\raisebox{2pt}{\protect\tikz{\protect\draw[-,red,solid,line width = 0.9pt](0,0) -- (5mm,0);}}}

\newcommand{\blackdotline}{\raisebox{2pt}{\protect\tikz{\protect\draw[-,black,dashed,line width = 0.9pt](0,0) -- (5mm,0);}}}

\newcommand{\redcircle}{\raisebox{0pt}{\protect\tikz{\protect\filldraw[fill=red, draw=red](2.5mm,0) circle [radius=1mm];}}}

\newcommand{\bluewithcircle}{\raisebox{0pt}{\protect\tikz{\protect\filldraw[fill=white, draw=blue](2.5mm,0) circle [radius=0.8mm];\protect\draw[-,blue,solid,line width = 1.0pt](0,0) -- (1.7mm,0);\protect\draw[-,blue,solid,line width = 1.0pt](3.3mm,0) -- (5mm,0);}}}

\newcommand{\greentriiangle}{\raisebox{0pt}{\protect\tikz{\protect\filldraw[fill=green, draw=green](2.5mm,0) (0.15,-0.1) -- (0.35,-0.1) -- (0.25,0.1)-- (0.15,-0.1);}}}
\newcommand{\bluetriiangle}{\raisebox{0pt}{\protect\tikz{\protect\filldraw[fill=blue, draw=blue](2.5mm,0) (0.15,-0.1) -- (0.35,-0.1) -- (0.25,0.1)-- (0.15,-0.1);}}}

\newcommand{\blacktriiangle}{\raisebox{0pt}{\protect\tikz{\protect\filldraw[fill=black, draw=black](2.5mm,0) (0.15,-0.1) -- (0.35,-0.1) -- (0.25,0.1)-- (0.15,-0.1);}}}
\newcommand{\megentasidetriiangle}{\raisebox{0pt}{\protect\tikz{\protect\filldraw[fill=magenta, draw=magenta, rotate=45](2.5mm,0) (0.15,-0.1) -- (0.35,-0.1) -- (0.25,0.1)-- (0.15,-0.1);}}}
\newcommand{\blueinvtriangle}{\raisebox{0pt}{\protect\tikz{\protect\filldraw[fill=blue, draw=blue](2.5mm,0) (0.15,-0.1) -- (0.35,-0.1) -- (0.25,-0.3)-- (0.15,-0.1);}}}
\newcommand{\magentainvtriangle}{\raisebox{0pt}{\protect\tikz{\protect\filldraw[fill=magenta, draw=magenta](2.5mm,0) (0.15,-0.1) -- (0.35,-0.1) -- (0.25,-0.3)-- (0.15,-0.1);}}}
\newcommand{\redinvtriangle}{\raisebox{0pt}{\protect\tikz{\protect\filldraw[fill=red, draw=red](2.5mm,0) (0.15,-0.1) -- (0.35,-0.1) -- (0.25,-0.3)-- (0.15,-0.1);}}}
\newcommand{\blacktriianglen}{\raisebox{0pt}{\protect\tikz{\protect\filldraw[fill=white, draw=black](2.5mm,0) (0.15,-0.1) -- (0.35,-0.1) -- (0.25,0.1)-- (0.15,-0.1); ;\protect\draw[black,solid](0,0) -- (5mm,0);}}}

\newcommand{\redwithsquare}{\raisebox{0pt}{\protect\tikz{\protect\filldraw[fill=white, draw=red](1.75mm, -0.75mm) rectangle ++(4.5pt,4.5pt);\protect\draw[red,solid](0,0) -- (5mm,0);}}}

\newcommand{\blacksequare}{\raisebox{0pt}{\protect\tikz{\protect\filldraw[fill=black, draw=black](1.75mm, -0.75mm) rectangle ++(4.5pt,4.5pt);}}}

\newcommand{\cyandiamond}{\raisebox{0pt}{\protect\tikz{\protect\filldraw[fill=cyan, draw=cyan, rotate=45](1.75mm, 0.75mm) rectangle ++(4.5pt,4.5pt);}}}

\newcommand{\olivediamond}{\raisebox{0pt}{\protect\tikz{\protect\filldraw[fill=green, draw=olive, rotate=45](1.75mm, 0.75mm) rectangle ++(4.5pt,4.5pt);}}}

\author{Ankit Patidar}
\author{Gaurav Goel}
\affiliation{Department of Chemical Engineering, Indian Institute of Technology Delhi, New Delhi, India}
\email{goelg@chemical.iitd.ac.in}

\title[An \textsf{achemso} demo]
  {MARTINI Coarse-grained Force Field for Thermoplastic Starch Nanocomposites}

\abbreviations{AA, CG, TMA, MMT, TPS, $T_g$, $R_g$, MSD}

\usepackage{setspace}
\begin{document}

\begin{tocentry}

{\centering
}
\end{tocentry}
\mciteErrorOnUnknownfalse
\doublespacing
\begin{abstract}
Thermoplastic starch (TPS) is an excellent film-forming material, and the addition of fillers such as tetramethylammonium-montmorillonite (TMA-MMT) clay has significantly expanded its use in packaging applications. We first used all-atom (AA) simulation to predict several macroscopic (Young's modulus, glass transition temperature, density) and microscopic (conformation along 1--4 and 1--6 glycosidic linkages, composite morphology) properties of TPS melt and TPS--TMA-MMT composite. The interplay of polymer-surface (weakly repulsive), plasticizer-surface (attractive), and polymer-plasticizer (weakly attractive) interactions lead to conformational and dynamics properties distinct from those in systems with either attractive or repulsive polymer-surface interactions.
A subset of AA properties was used to parameterize the MARTINI coarse-grained (CG) force field (FF) for the melt and composite systems. Specifically, we determined the missing bonded parameters of amylose and amylopectin and rationalized the bead types for 1--4 and 1--6 linked $\alpha$-D glucose using two-body excess entropy, density, and bond and angle distributions in AA TPS melt. The MARTINI CG model for TPS was combined with an existing parameter set for TMA-MMT. The liquid-liquid partitioning-based MARTINI-2 FF shows freezing and compaction of polymer chains near the sheet surface, further accentuated by lowering of dispersive interactions between pairs of high covalent-coordination ring units of TPS polymers and MMT sheet. A rescaling of the effective dispersive component of TPS-MMT cross-interactions was used to optimize the FF for the composite system, with structural (chain size distribution), thermodynamic (chain conformational entropy, density), and dynamic (self-diffusion coefficient) properties obtained from long AA simulations forming the constraints for optimization. The obtained CG FF parameters provided excellent estimates for several other properties of the melt and composite systems not used in parameter estimation, thus establishing the robustness of the developed model. 
\end{abstract}
\maketitle

\section{Introduction}
\begin{figure*}[!htb]
     \centering
     \includegraphics[width=14cm,height=10cm]{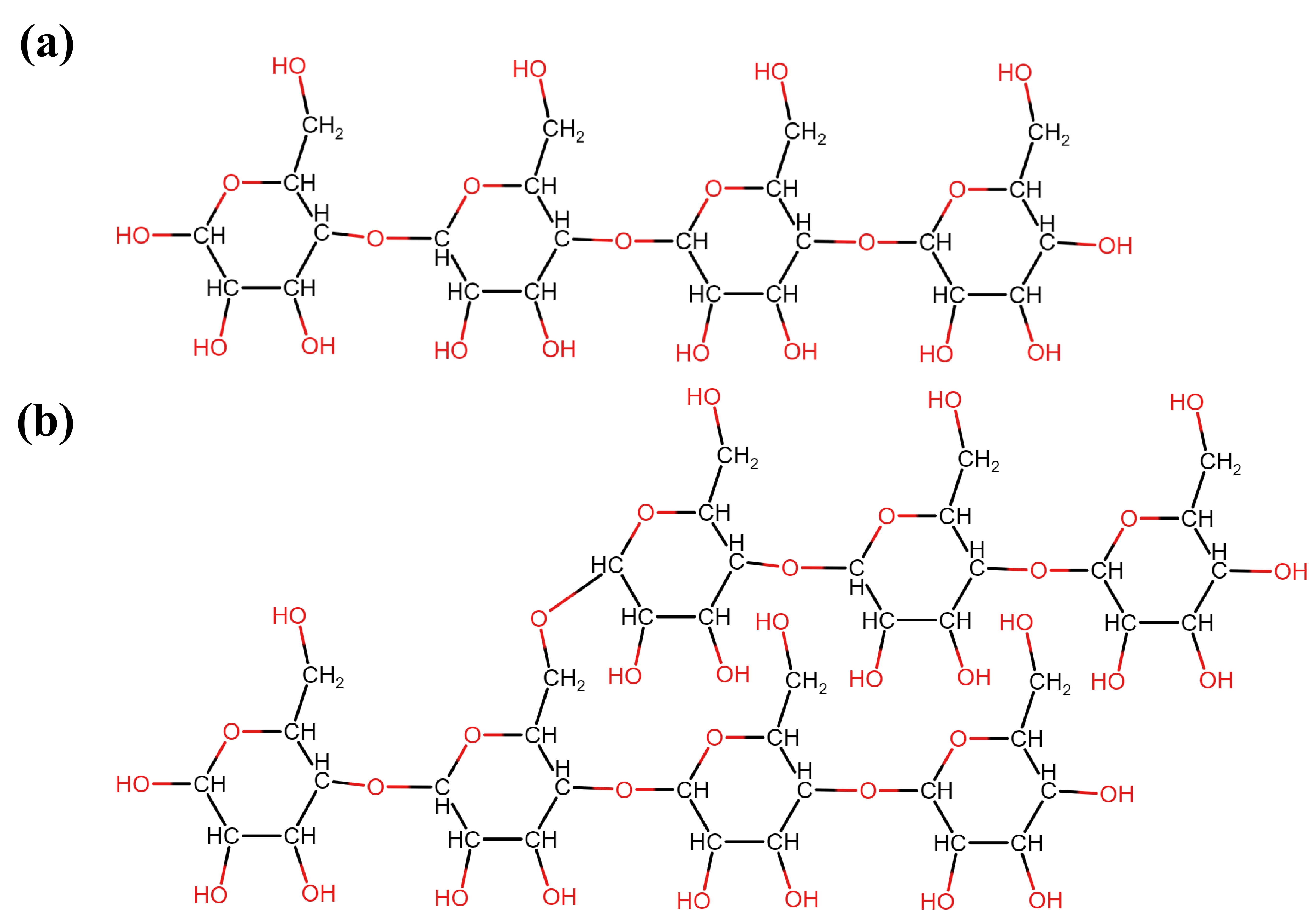}
     \caption{\textbf{Chemical structure of amylose and amylopectin }. A chemical representation of \textbf{(a)} amylose and \textbf{(b)} amylopectin chain consisting of four and seven $\alpha$-D glucose monomers, respectively.}
     \label{fig:structure}
 \end{figure*}

 Thermoplastic starch (TPS) is obtained by gelatinization of native starch, a polysaccharide consisting of amylose and amylopectin, by addition of plasticizers, such as glycerol, sorbitol, water, etc. \cite{carvalho2008starch}. Amylose is a linear polymer of 1-4 linked $\alpha$-D glucose and amylopectin has additional 1-6 branching on the 1-4 backbone (Figure \ref{fig:structure}). Low cost, excellent film-forming properties (low glass transition temperature, easy melt processability, acceptable elongation at break) \cite{jimenez2012edible,attaran2017materials}, low oxygen permeability, and biodegradability \cite{babu2013current} make TPS an attractive option for food packaging, drug delivery, paper additive, etc. \cite{khan2017thermoplastic, stepto2003processing, de2022thermoplastic,kaboorani2021tailoring, correa2022biodegradability}. However, its poor mechanical properties (low Young's modulus and tensile strength, limited elongation at break) \cite{Cyras2008,chocyk2015influence,domene2019influence} and low moisture resistance \cite{salaberria2015role} necessitate the addition of components with complementary properties, such as addition of nanofillers \cite{Rezaei2015, zhou2015starch, Suter2015}, blending with other polymers \cite{altayan2022toward,baumberger1998use}. Several studies have shown improvement in tensile strength and moisture barrier of TPS films on adding surface polarity-matched phyllosilicate clays, such as organically-modified montmorillonite (MMT) \cite{park2002preparation,zhou2015starch, schlemmer2010morphological,Cyras2008, majdzadeh2010optimization,rivadeneira2021green,Huang2004}. In general, wide availability, low cost, and surface tunability make MMT a particularly suitable nanofiller for modulating properties of polymeric materials \cite{ Rezaei2015,Krishnamoorti1996,alias2021hybridization}. For example, blending with polylactic acid (PLA) was shown to nominally reduced water sorption of a TPS film, but microphase separation of two polymer phases led to poorer mechanical properties \cite{ayana2014highly}. Here, addition of \SI{1}{\percent} Na-MMT) resulted in a lower number-average domain diameter of PLA (from \SI{21.3}{\micro\meter} to \SI{13.45}{\micro\meter}) and an improved tensile strength (from \SI{5.63}{\mega\pascal} to \SI{7.95}{\mega\pascal}). The properties of composite materials depend on various factors, including polymer chemistry, polymer-filler interactions, and filler aspect ratio, which consequently determine filler particle dispersion and composite's morphology \cite{Mao2012,Huang2004,Rezaei2015,Suter2015}. The identification of specific nanoparticles and other topological parameters for improving properties of interest and optimization in the composition-property space requires a time- and resource-intensive exploration \cite{Mao2012,Huang2004,Rezaei2015,Suter2015}, made more challenging for starch-based composites due to a large diversity in the base components (TPS is a multi-component material) and chemical heterogeneity present in each component (e.g., primary structure diversity of amylopectin). Molecular dynamics (MD) simulations provide a viable first step to characterize structure-property relationships of multi-component systems and obtain narrow bounds on the composition space for a target property, thus complementing efforts to develop new polymer composites \cite{huang2022molecular, guo2023molecular, lin2023multiscale, gartner2019modeling}. MD simulations using all-atom (AA) representation of polysaccharides \cite{guvench2008additive,hatcher2009charmm,geronimo2019kinetic,kirschner2008glycam06,plazinski2016revision} have been shown to accurately predict a variety of properties of some simple systems, for example, glass transition temperature and Young's modulus of TPS melt \cite{Ozeren2020,Ozeren2020a}, chain size distribution of amylose in water \cite{Koneru2019}, and crystalline packing of cellulose fibers \cite{mazeau2003molecular}. However, slow relaxation dynamics and disparate length scales of the components preclude the applicability of AA simulations to several systems of fundamental and technological interest. For example, the characteristic relaxation time of a small 20-unit amylose (\SI{3260.86}{\gram\per\mol}) in its melt was reported to be \SI{404.7}{\nano\second} at \SI{700}{\kelvin} (much above typical processing temperature) \cite{gatsiou2022atomistic}, eight orders of magnitude higher than the AA time step even for this limiting case. The presence of high molecular weight polymer chains (radius of gyration $\sim \SI{50}{\nano\meter}$) and highly anisotropic fillers such as $\sim \SI{200}{\nano\meter}$ diameter bentonite clay) bentonite clay in TPS film make property determination using AA simulations computationally intractable. The coarse-grained (CG) force fields, wherein some of the AA degrees of freedom are integrated out, can be $\sim$ 100-fold faster than AA simulations even at the lower-end of coarse-graining. CG approaches employed to study polymer composites at some level of relevant length- and timescales can be broadly divided into two categories based on the treatment of interaction between various components. In one case, the enthalpic contribution is modeled by a minimal, single-parameter effective interaction between large units (a Kuhn monomer of $\sim$ \SI{1}{\nano\meter}, a nanoparticle, etc.). The effective interaction parameters are then used in mesoscale dissipative particle dynamics (DPD) simulations \cite{Groot1997,Singh2018,Fu2013,Scocchi2007,long2006nonlinear,sliozberg2020dissipative,khani2015polymer,wang2021dissipative,ju2013miscibility} or field-based / integral-equation theories \cite{schweizer1992reference, schweizer1997polymer, zirkel2002small, wu2006density,oxtoby2002density,helfand1975theory,helfand1976block,lowden1973solution,sung2005integral, hall2011impact,martin2016using, hsu1979rism,martin2018pyprism}, to characterize structure and thermodynamics of diverse polymeric systems. In a multiscale DPD scheme, integral equation theory was used to obtain an effective softcore interaction between polymer chains, enabling use of a very large time step (\SI{}{\pico\second} - \SI{}{\nano\second}) \cite{Clark2010,Clark2012}, and its application to polyethylene melt \cite{Lyubimov2010, Lyubimov2013} and blends \cite{Mccarty2010} showed a fairly good comparison with structural (radius of gyration and radial distribution function) and dynamic properties (diffusion coefficient of chain center of mass) obtained in united atom (UA) simulations. These methods, based on a simplified model for enthalpy, have provided qualitative and, in several cases, quantitatively accurate predictions for the phase diagram and the equilibrium properties of large-scale polymer systems \cite{wang2021dissipative,karatrantos2016modeling,zeng2008multiscale,gooneie2017review,glotzer2002molecular}. The interaction parameter is dependent on the thermodynamic state variables (temperature, pressure, composition) and needs to be obtained semi-empirically from molecular simulations, liquid-state theories, or experiments at a sufficiently large number of state points, wherein an increase in system heterogeneity significantly expands the phase space for interaction parameter determination.\\

Chemical-specific CG models provide an alternative with possible state-point transferability and good accuracy across highly diverse systems. Herein, 4-10 atoms are grouped in a single bead, with effective pair interaction obtained from a combination of AA simulations and experimental data. This mapping from AA to CG can be obtained by using structure distribution functions as constraints, as done in iterative Boltzmann inversion (IBI) \cite{reith2001mapping,reith2003deriving} and force matching \cite{izvekov2005multiscale} or by using thermodynamic properties as constraints, such as equilibrium partitioning between polar and apolar solvents used for MARTINI CG force field \cite{Marrink2007}. In structure-based methods, parameters need to be optimized using data from multiple state points \cite{moore2014derivation}, and can still exhibit environmental dependence.
The MARTINI force field has been demonstrated to mitigate these limitations. For instance, the MARTINI CG parameters for polypropylene were shown to accurately reproduce (values obtained in AA simulation) the radius of gyration in various chemical environments (good, bad, and theta solvent conditions) as well as the partitioning of polyethylene-polypropylene blends near phospholipid membranes \cite{panizon2015martini}. However, a reduction in self- (saccharides molecules) \cite{Schmalhorst2017,shivgan2020extending} or cross-interaction (graphene-small organic molecules) \cite{wu2012coarse,gobbo2013martini,piskorz2019nucleation} was required in some cases, highlighting a need for further parameter optimization.

Given the wide interest in phyllosilicate-based composites, we have developed MARTINI CG parameters for a TPS--[tetra methyl ammonium(TMA)]-MMT system in this study. The CG parameters for some individual TPS components (amylose \cite{Lopez2009}, sorbitol \cite{sukenik2015osmolyte}, water \cite{michalowsky2017refined}) and TMA-MMT \cite{Khan2019} have been shown to reproduce several properties determined from AA simulations. However, to the best of our knowledge, MARTINI CG FF for a TPS melt or its composites is not available. Here, we have re-parameterized and developed missing bonded parameters for the CG model of two polymeric components (amylose and amylopectin) by using data from AA simulation of TPS melt, rationalized the bead type assignment for 1-4 and 1-6 linked $\alpha$-D glucose, and determined the cross-interaction levels for all TPS and TMA-MMT beads. The developed parameter set was extensively validated by comparing structural (chain size distribution), thermodynamic (chain conformational entropy, density, excess entropy), and dynamic (self-diffusion coefficient) properties against sufficiently long AA simulations. Finally, predictions for radial distribution function (RDF) and two-body excess entropy in the composite system, two properties not used in CG parameterization, were used to test the robustness of developed CG parameters.
\section{Methods}
\subsection{System and Simulation Details}

\subsubsection{AA MD Simulations}
The initial structures and CHARMM36 \cite{guvench2009charmm,guvench7ii} parameters of amylopectin (thirty-six monomers), amylose (eighteen monomers), and sorbitol were obtained using CHARMM-GUI \cite{jo2008charmm, guvench7ii, guvench2009charmm}. We used a TIP3P water model and the INTERFACE force field \cite{heinz2005force} for the TMA-MMT sheet. A low density ($\rho = \SI{117.62}{\kilo\gram\per\cubic\meter}$) TPS configuration (55327 atoms) consisting of (weight percent) \SI{50}{\percent} amylopectin, \SI{20}{\percent} amylose, \SI{29}{\percent} sorbitol, and \SI{1}{\percent} water was prepared in Packmol \cite{martinez2009packmol}. A high-density system was obtained by rescaling the box in multiple steps followed by temperature-pressure ($T$-$P$) annealing cycles (details in Section \ref{annealing}). Densities of \SI{1227.75} and \SI{1227.62}{\kilo\gram\per\cubic\meter} at \SI{613}{\kelvin}, \SI{1}{\bar} were obtained after the first two $T$-$P$ cycles, respectively, indicating the adequacy of only one cycle. The obtained value compares well with a TPS melt density of \SI{1369.80}{\kilo\gram\per\cubic\meter} from AA simulation at \SI{550}{\kelvin}, \SI{1}{\bar} \cite{Ozeren2020} and \SI{1442}{\kilo\gram\per\cubic\meter} at \SI{300}{\kelvin}, \SI{1}{\bar} from experiments \cite{Lyckfeldt1998}. For composite simulations, a $\SI{10.32}{\nano\meter} \times \SI{10.71}{\nano\meter} \times \SI{1.2}{\nano\meter}$ TMA-MMT sheet, connected across $x$-$y$ periodic boundaries to model an infinitely large sheet, was placed between two equilibrated TPS melt boxes (Figure \ref{fig:AACGcomposite} (a)). An \SI{82.5}{\percent} TPS and \SI{17.5}{\percent} TMA-MMT system with dimensions \SI{10.32}{\nano\meter}$\times$\SI{10.71}{\nano\meter}$\times$\SI{11.40}{\nano\meter} was obtained after one $T$-$P$ annealing cycle. All melt and composite properties were calculated from a \SI{1100}{\nano\second} MD simulation started from the configuration at the end of the first $T$-$P$ annealing cycle. MD simulations were performed using GROMACS 2021.4 \cite{abraham2015gromacs}, with a time step of \SI{1}{\femto\second}, V-rescale thermostat ($\tau$=\SI{1}{\pico\second}) \cite{bussi2007canonical}, and either the Berendsen barostat ($\tau$=\SI{2}{\pico\second}, first \SI{100}{\nano\second}) \cite{berendsen1984molecular} or the Parrinello-Rahman barostat ($\tau$=\SI{5}{\pico\second}) \cite{parrinello1981polymorphic}.
   
\subsubsection{CG MD Simulations} 
\begin{figure}[!htb]
     \centering
     \includegraphics[width=8.25cm,height=10.25cm]{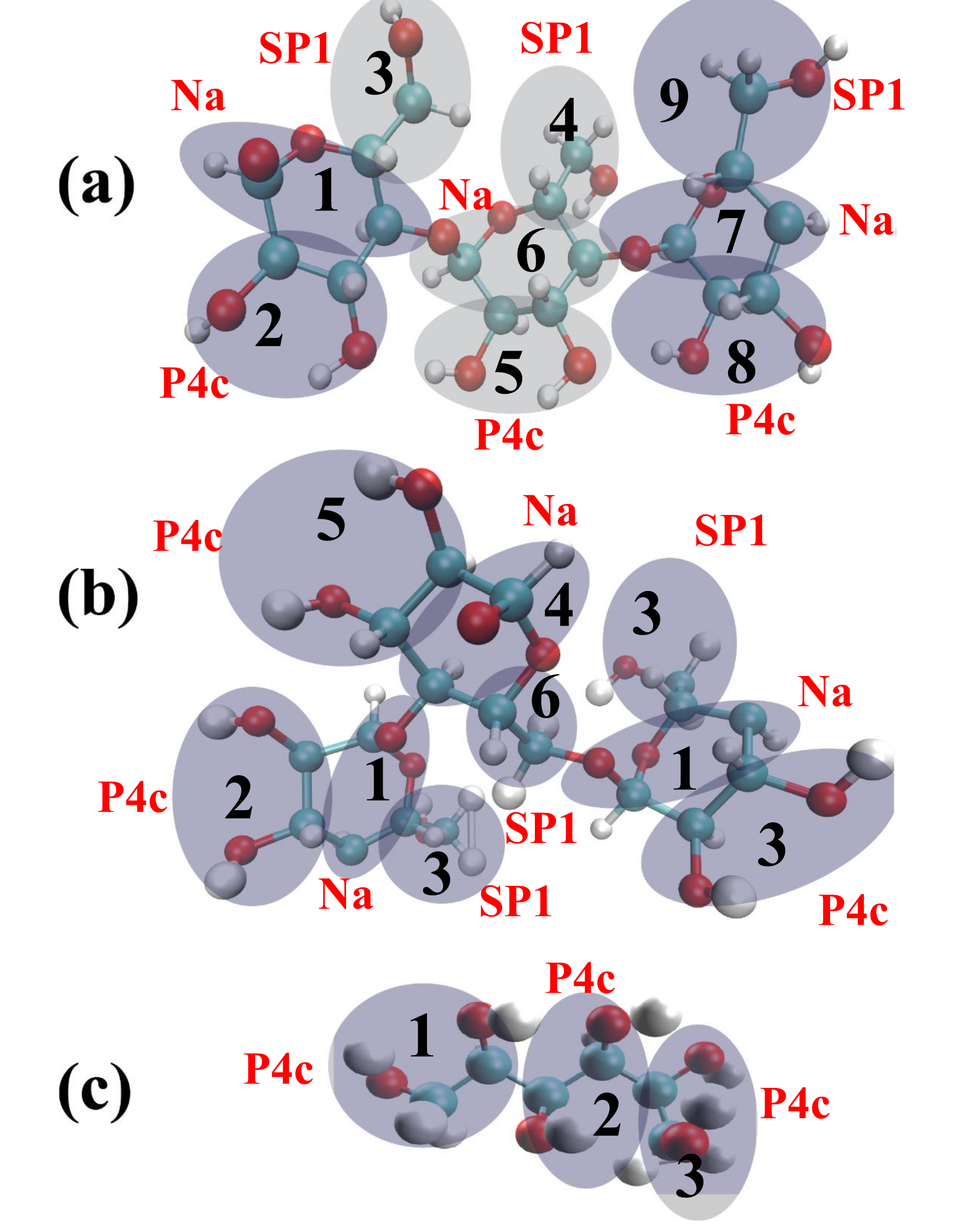}
     \caption{\textbf{Mapping scheme for TPS components}. \textbf{(a)} A trimeric 1--4 bonded $\alpha$-D glucose oligomer (represents terminal and middle monomer of amylose), \textbf{(b)} A trimeric 1--4 or 1--6 bonded $\alpha$-D glucose oligomer (represents the branch point in amylopectin), and \textbf{(c)} sorbitol. Red, white, and cyan represent oxygen, hydrogen, and carbon atoms, respectively, with atoms in a single CG bead grouped within shaded circles. The bead number and type are indicated by black and red color labels, respectively.}
     \label{fig:mapping}
 \end{figure}

\begin{figure*}[!htb]
     \centering
     \includegraphics[width=15cm,height=10cm]{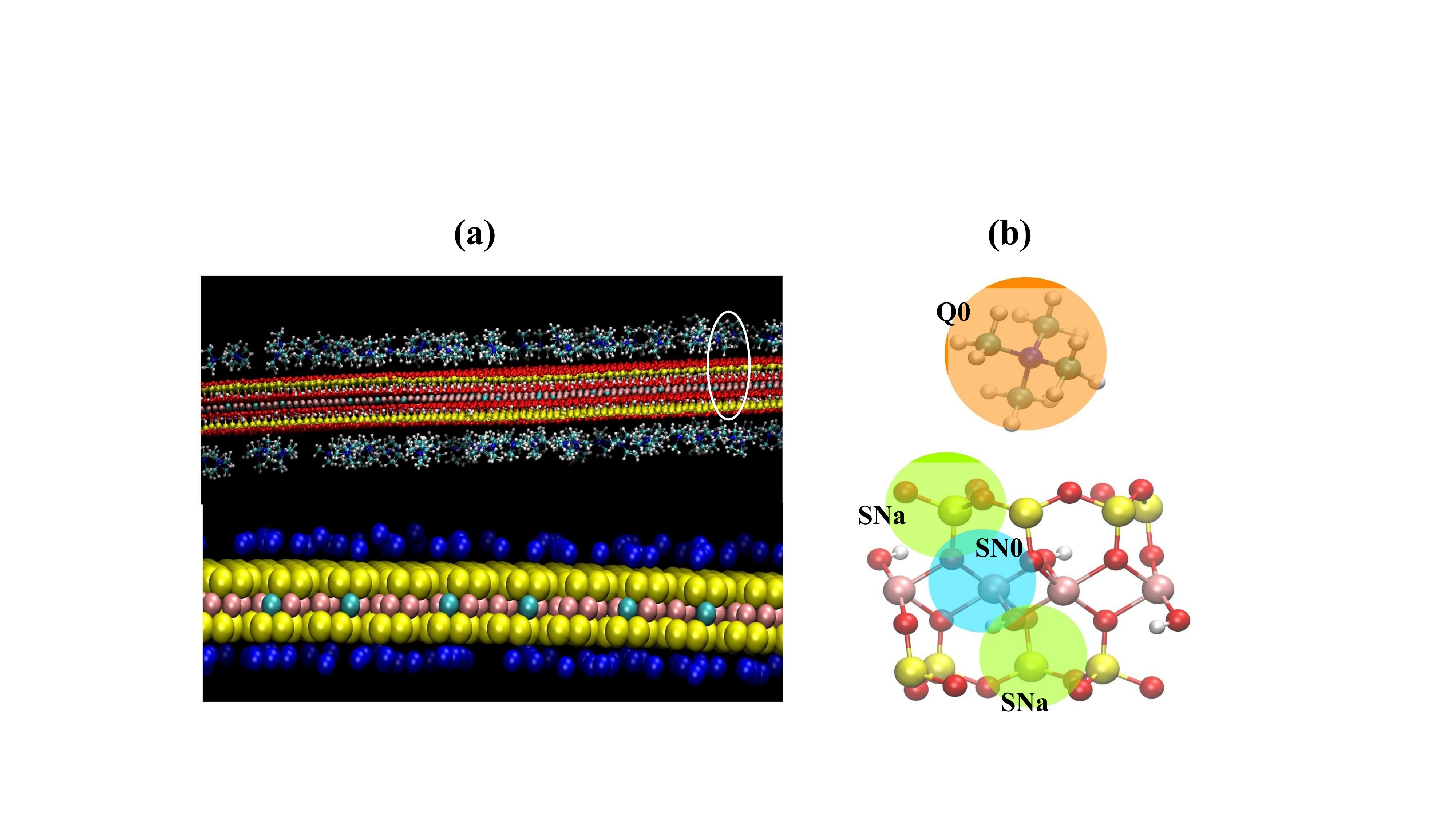}
     \caption{\textbf{AA and CG representation of TMA-MMT sheet with the Mapping scheme}. \textbf{(a)} shows the AA and CG representation of the TMA-MMT sheet. \textbf{(b)} shows the mapping scheme and MARTINI bead type of the highlighted portion of the sheet in (a)}
     \label{fig:sheetmapping}
 \end{figure*}

The initial CG structures for TPS melt and TPS--TMA-MMT composite were obtained from the last frame of the corresponding \SI{1100}{\nano\second} AA simulation by placing CG beads at the center of mass positions (Equation \ref{eq:mapping}) of constituent atoms (mapping scheme in Figure \ref{fig:mapping}). Force field parameters at the level of MARTINI v2.2\cite{Marrink2007} were either directly used, if available and found suitable, or were obtained using reference AA simulations. For all CG simulations, a time step of \SI{5}{\femto\second} was used, the system was coupled to the V-rescale thermostat ($\tau$=\SI{1}{\pico\second}) and either the Berendsen barostat ($\tau$=\SI{8}{\pico\second}, first \SI{100}{\nano\second}) or the Parrinello-Rahman barostat ($\tau$=\SI{12}{\pico\second}). The CG potential is given as per Equation \ref{eq:LJ}.

\begin{multline}\label{eq:LJ}
U=4\epsilon_{ij}\left[\left(\frac{\sigma_{ij}}{r_{ij}}\right)^{12}-\left(\frac{\sigma_{ij}}{r_{ij}}\right)^6\right]+\frac{q_{i}q_{j}}{4{\pi}\epsilon_0\epsilon_{rel}{r_{ij}}}\\
+\frac{1}{2}K_l(l_{ij}-l_0)^2+\frac{1}{2}K_\theta(cos\theta_{ijk}-cos\theta_0)^2\\+K_\phi[1+(cos\phi_{ijkl}-cos\phi_0)],
\end{multline}
where $\sigma_{ij}$, $\epsilon_{ij}$, $q$, and $\epsilon_{rel}$ are bead diameter, interaction strength, bead charge, and relative dielectric constant, respectively. The force constants for bond ($l_{ij}$), angle ($\theta_{ijk}$), and torsion ($\phi_{ijkl}$) potentials are $K_l$, $K_\theta$, and $K_\phi$, respectively. $\sigma_{ij}=\SI{0.47}{\nano\meter}$ for the regular beads (four heavy atoms) and \SI{0.43}{\nano\meter} for the small beads (s-type, 2-3 heavy atoms). The choice of $\epsilon_{ij}$ is linked to the functional group(s) in the CG bead, with primary categorization defined as polar (P), neutral (N), apolar (C), or charged (Q), and level within a category assigned by an integer in range 0--6 (e.g., P4 is more polar than P1, C4 is more apolar than C1). We used a P4c bead for three diol units of sorbitol (see Figure \ref{fig:mapping}), a CG model shown to be in qualitative agreement with experiments on the kinetics of aggregation of amyloid- and elastin-like peptides in presence of sorbitol \cite{sukenik2015osmolyte}. The refined polarizable model was used for water since the standard MARTINI water model (one P4 bead) exhibits unphysical freezing near surfaces \cite{yesylevskyy2010polarizable, michalowsky2017refined}. The parameters for the TMA-MMT sheet were taken from Khan and Goel \cite{Khan2019} who found several properties of hydrophobic polymers--TMA-MMT composites to be in a good agreement with AA simulations  (Supplementary Information Figure \ref{fig:sheetmapping}). The bonded parameters for 1–4 linked $\alpha$-D glucose were either taken from Lopez et al.'s study \cite{Lopez2009} on short amylose chains in polar (water) and non-polar (nonane) solvents (labeled CG1 model)  or obtained as best-fit models to the corresponding pair, triplet, and quadruplet distributions in AA simulations of TPS melt (labeled CG2 model). The bonded parameters for 1-6 linked $\alpha$-D glucose were similarly obtained from the AA simulations of TPS melt (for both CG1 and CG2).\\ 
 
\subsection{Property Calculation}

In the composite system, the near- and far-sheet regions were classified based on density profiles of TPS components along the normal to the TMA-MMT sheet ($z$-axis). 
For polymer chains, the $z$-component of the distance of the center of mass (COM) of a given trimeric segment ($\sim$ persistence length (\SI{0.86}{\nano\meter}) of amylose) from the sheet was used to classify it as a near or a far segment, with a chain labeled as a near/far-region chain if $>\SI{70}{\percent}$ segments were in the near/far region. Unless otherwise indicated, all properties are calculated at \SI{613}{\kelvin} and \SI{1}{\bar} using the last \SI{900}{\nano\second} trajectory (a \SI{200}{\nano\second} convergence time for some properties in total \SI{1100}{\nano\second} trajectory).

\subsubsection{Structural and Dynamic Properties}

Density, radial distribution function ($RDF$), persistence length (${l_p}$), and radius of gyration (${R_g}$) of TPS melt and composite systems were estimated using GROMACS utilities. COMs of atoms making the CG beads were used for $RDF$ calculation in the AA simulations to allow comparison with the CG simulations. An exclusion of up to four CG-bonded neighbors or up to 16 AA-bonded neighbors, equivalent to an average COM separation of \SI{1.29}{\nano\meter}, was used for $RDF$ calculation. For the composite system, a 2D $RDF$ was calculated in \SI{0.37}{\nano\meter} slices parallel (in the $x-y$ plane) to the TMA-MMT sheet. The data is reported as average over four slices in the near region and five slices in the far region. 

The diffusion coefficients of sorbitol and trimeric polymer chain segments (amylose and amylopectin) were calculated from the corresponding temporal profile of mean-square-displacement (${MSD} (t)$) using the Einstein's diffusion equation (Equation \ref{eq:MSD}). The AA and CG data were compared using the normalized diffusion coefficient, $D_\mathrm{n}$, defined as the ratio of lateral diffusion coefficient in the composite (in the $x-y$ plane) to the melt diffusion coefficient.

\subsubsection{Thermodynamic Properties}
The TPS melt was cooled from \SI{613}{\kelvin} to \SI{150}{\kelvin} in steps of \SI{25}{\kelvin}, wherein a \SI{4}{\nano\second} NPT simulation was performed at each step, with last \SI{500}{\pico\second} trajectory at each step used for density calculation. The intersection of linear fits to the temperature-density data in the rubbery (high temperature) and glassy (low temperature) regions was used as an estimate for $T_\mathrm{g,sim}$.  The Williams-Landel-Ferry (WLF) equation \cite{williams1955temperature} (Equation \ref{eq:WLF}) was used to obtain an estimate for the glass transition temperature at a typical experimental cooling rate of \SI{10}{\kelvin\per\minute} ($T_\mathrm{g,exp}$) from the simulation estimate $T_\mathrm{g,sim}$ obtained at \SI{3.75e11}{\kelvin\per\minute}.

\begin{equation}\label{eq:WLF}
\mathrm{log_{10}} \frac{\tau_{g,\mathrm{sim}}}{\tau_{g,\mathrm{exp}}} =\frac{\mathrm{-C_1*(T}_{g,\mathrm{sim}} -\mathrm{T}_{g,\mathrm{exp}})}{\mathrm{C_2 + T}_{g,\mathrm{sim}}-\mathrm{T}_{g,\mathrm{exp}}},
\end{equation}

 where $C_1=17.44$ and $C_2=\SI{51.6}{\kelvin}$ (constant for several polymers), $\tau_\mathrm{g,sim}$ and $\tau_\mathrm{g,exp}$ are the inverse of cooling rates in simulation and experiment, respectively.\\
The pairwise ($S_{2,\alpha\beta}$) and total ($S_\mathrm{2}$) two-body excess entropy was calculated using the corresponding $RDFs$ ($g_{\alpha\beta}$) as per Equations \ref{eqn:tbe} and \ref{eq:overalltbe} (integral calculated up to \SI{2.5}{\nano\meter}), respectively \cite{sharma2008estimating}. 

\begin{multline}
   \mathrm{S_{2,\alpha\beta}}=-2\pi\rho\int_{0}^{\infty}g_{\alpha\beta}(r)\mathrm{ln}\,g_{\alpha\beta} (r)-\\
   [g_{\alpha\beta}(r)-1]r^2dr
  \label{eqn:tbe}
\end{multline}
\begin{equation}\label{eq:overalltbe}
\mathrm{S_{2}}=\sum_{\alpha,\beta}{x_\alpha x_\beta}\mathrm{S_{\alpha\beta}},
\end{equation}
where $x_{\alpha}$ and $x_{\beta}$ are the mole fractions of $\alpha$ and $\beta$ components, respectively. Equation \ref{eqn:tbe} was suitably modified to integrate 2D RDFs for the composite system over cylindrical disks (Equation \ref{eqn:tbecomp}).

The configurational entropy ($S_\mathrm{c}$) of amylose and amylopectin chains was calculated from the covariance matrix of each chain (after removing rigid-body translational and rotational motion) using the Schlitter's equation \cite{schlitter1993estimation}, implemented with GROMACS utility anaeig. 

\subsubsection{Mechanical properties}

Young's modulus of TPS melt and TMA-MMT sheet was calculated from AA simulations at \SI{300}{\kelvin} and \SI{1}{\bar}. A uniaxial strain along $x$-axis was added in steps of \SI{0.1}{\percent} of the initial box length, with a \SI{2}{\nano\second} $NPT$ simulation at each step (anisotropic pressure coupling for $y$ and $z$). The stress at every step was calculated from the ${Pxx}$ Virial component averaged over the last \SI{300}{\pico\second} of the trajectory, and the slope of the stress-strain curve in the linear regime was used to estimate the Young's modulus. Maximum strain value ($\sigma_\mathrm{max}$) marks the onset of the plateau region in the stress-strain profile.

\section{Results and discussion}

\subsection{Validation of TPS AA parameters}

\begin{figure}[!htb]
     \centering
     \includegraphics[width=8.25cm,height=4.25cm]{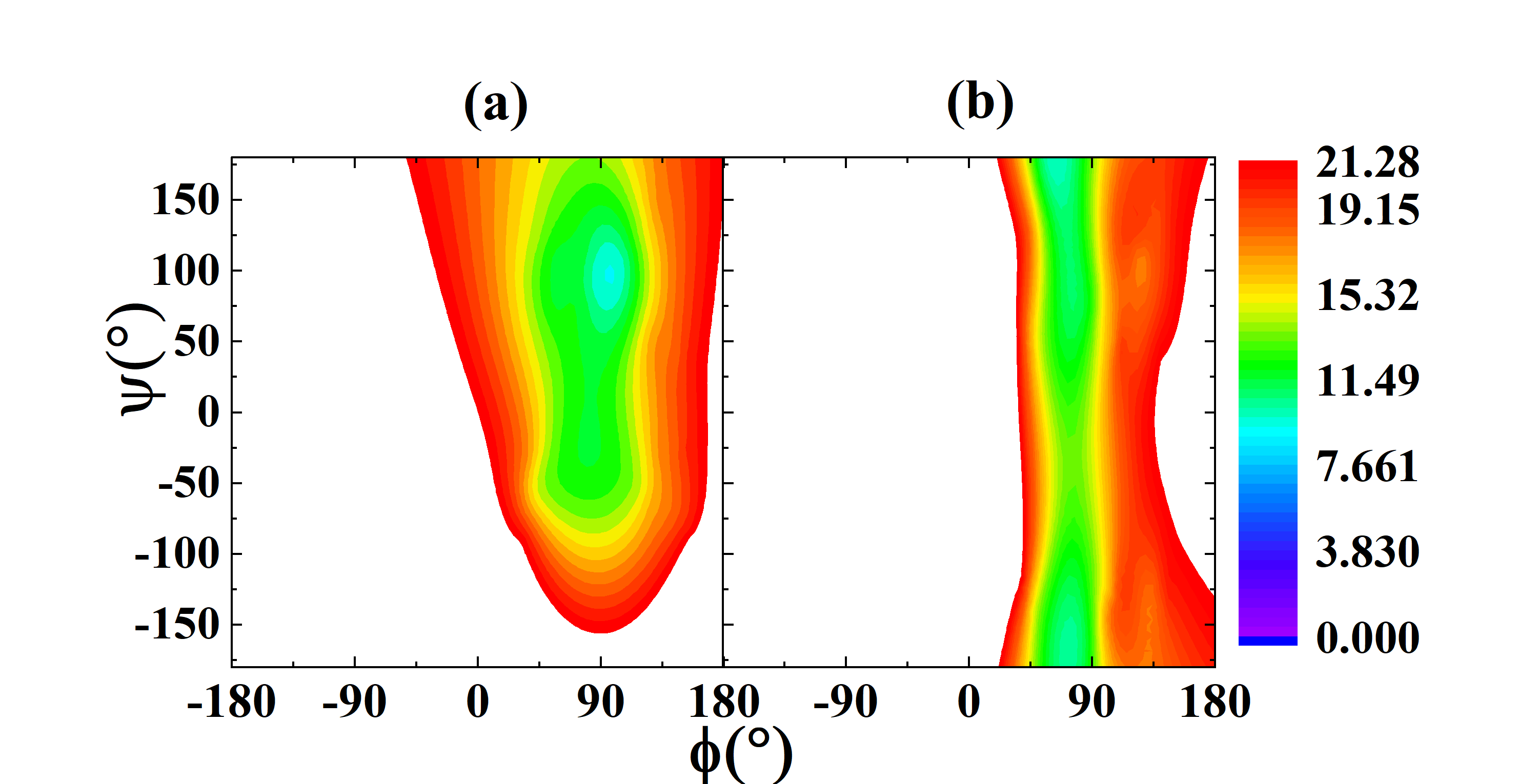}
     \caption{\textbf{Polymer chain conformation at inter-ring links in a TPS melt}. $\phi$-$\psi$ torsion angle distribution in TPS melt for \textbf{(a)} 1--4 and \textbf{(b)} 1--6 glycosidic linkages, shown as the potential of mean force (PMF) profile (color scale in \SI{}{\kilo\joule\per\mole}) obtained from the probability distributions averaged over \SI{900}{\nano\second} AA simulation at \SI{613}{\kelvin} and \SI{1}{\bar}.}
     \label{fig:1416diah}
\end{figure}
 
\begin{figure*}[!htb]
     \centering
     \includegraphics[width=10cm,height=10cm]{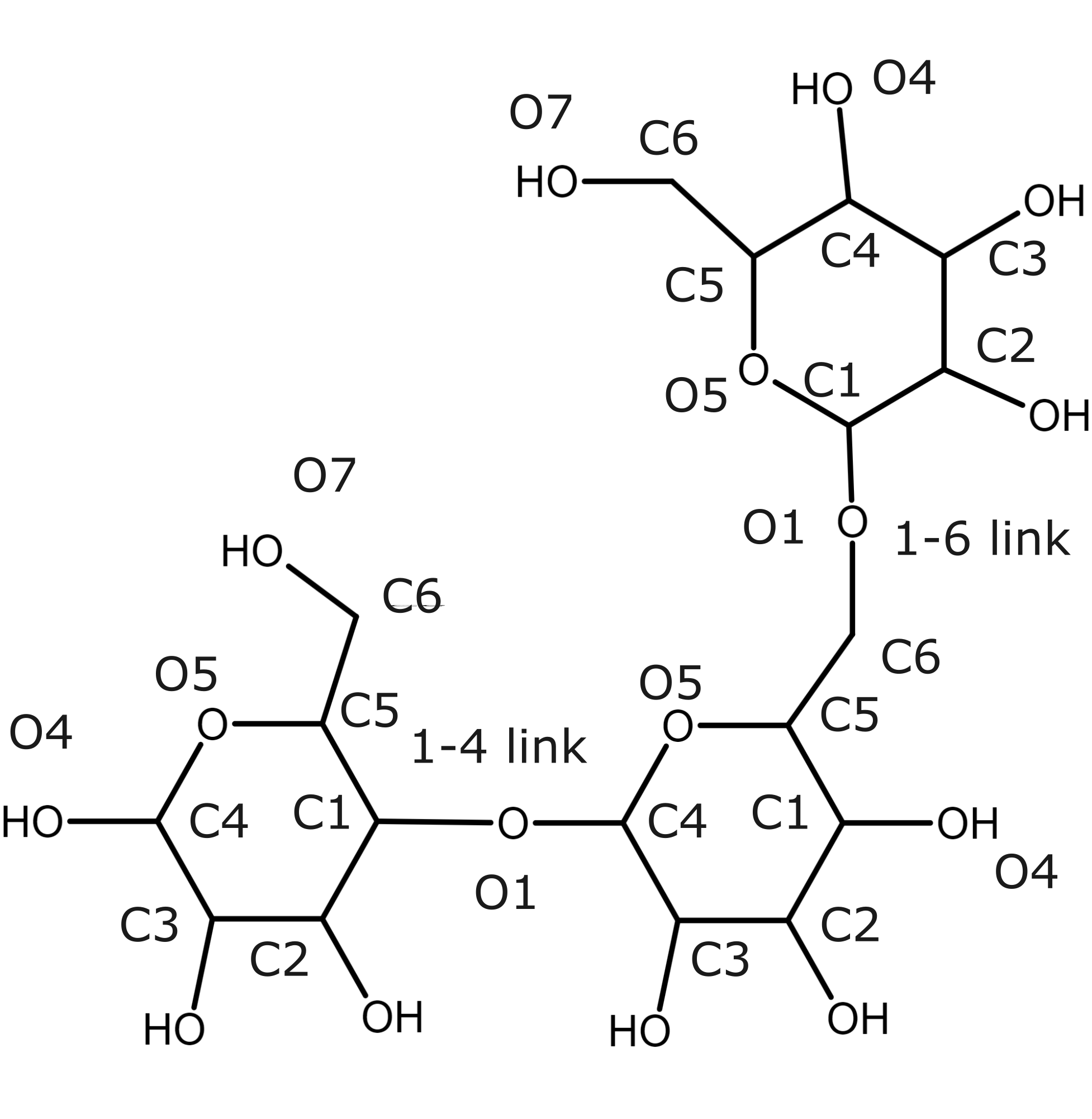}
     \caption{\textbf{AA representation of 1--4 and 1--6 linkage }. The 1--4 $\alpha$-D glucose linkage is represented by $\phi=\angle\mathrm{O_5}-\mathrm{C_4}-\mathrm{O_1}-\mathrm{C_1}$, $\psi=\angle\mathrm{C_1}-\mathrm{O_1}-\mathrm{C_4}-\mathrm{C_3}$, and 1--6 $\alpha$-D glucose linkage is represented by $\phi=\angle\mathrm{O_5}-\mathrm{C_1}-\mathrm{O_1}-\mathrm{C_6}$, $\psi=\angle\mathrm{C_1}-\mathrm{O_1}-\mathrm{C_6}-\mathrm{C_5}$.}
     \label{fig:psiphifig}
 \end{figure*}

The local chain conformation in TPS melt was characterized using the torsion angles for 1--4 and 1--6 glycosidic linkages, viz. $\phi=\angle{\mathrm{O_5-C_4-O_1-C_1}}$ and $\psi=\angle{\mathrm{C_1-O_1-C_4-C_3}}$ for 1--4, and $\phi=\angle{\mathrm{O_5-C_1-O_1-C_6}}$ and $\psi=\angle{\mathrm{C_1-O_1-C_6-C_5}}$ for 1--6 (Figure \ref{fig:psiphifig}). Figure \ref{fig:1416diah} (a) shows syn- and anti-states as two dominant conformers for the 1--4 linkage, with peak at ($\phi$,$\psi$) equal to (\SI{90}{\degree},\SI{100}{\degree}) and (\SI{100}{\degree},\SI{-50}{\degree}), respectively. This agrees well with the AA simulations of an amylose oligomer in water using the same force field ($syn$: (\SI{90}{\degree},\SI{100}{\degree}) and $anti$: (\SI{80}{\degree},\SI{-50}{\degree})) \cite{Koneru2019} and XRD spectroscopy of a hydrated cycloamylose ($syn$: (\SI{103.6}{\degree},\SI{115.1}{\degree}) and $anti$: (\SI{88.1}{\degree},\SI{-48.4}{\degree})) \cite{Gessler1999}.
The conformational space of 1--6 linkage has a single peak at (\SI{70}{\degree},\SI{180}{\degree}) (Figure \ref{fig:1416diah}(b)), in close agreement with replica exchange simulation and NMR spectroscopy on $\alpha$-1--6 glucopyranoside dimer (($\phi$,$\psi=$)(\SI{70}{\degree},\SI{180}{\degree})) \cite{Patel2014}.\\

  \begin{figure}[!htb]
     \centering
     \includegraphics[width=8.25cm,height=9.0cm]{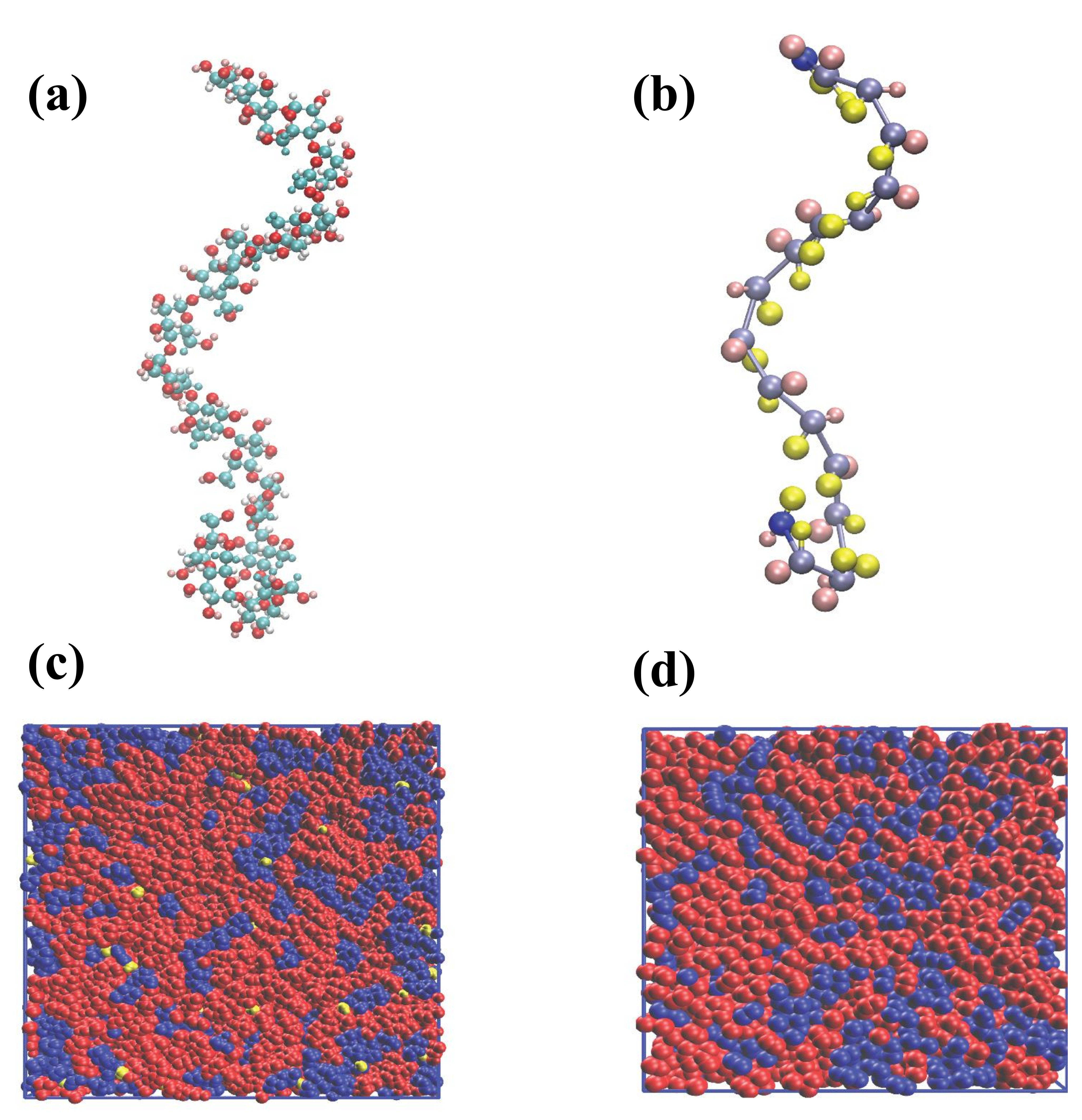}
     \caption{\textbf{AA and CG representation of TPS melt}.  A 18-mer amylose chain in \textbf{(a)} AA (oxygen, hydrogen, and carbon atoms in red, white, and cyan, respectively) and \textbf{(b)} CG (P1, SP1, P4c, and Na beads in blue, pink, yellow, and ice blue, respectively) representations. \textbf{(c)} and \textbf{(d)} TPS melt simulation box obtained at end of \SI{1100}{\nano\second} AA simulation at \SI{613}{\kelvin} and \SI{1}{\bar} in AA and CG, respectively. Polymer, sorbitol, and water beads in red, blue, and yellow, respectively.}
     \label{fig:AACGmelt}
 \end{figure}

\begin{table}[!htb]
\begin{minipage}{0.5\textwidth}
  \centering
  \renewcommand{\thempfootnote}{\alph{mpfootnote}}
  \renewcommand{\arraystretch}{0.7}
  \setlength{\tabcolsep}{2pt}
  \captionof{table}{\textbf{A comparison of various properties of TPS melt obtained from AA simulations with previous simulations and experimental measurements.}} 
  \centering
  \label{tab:AAproperties}
  \begin{tabular}{l  c c c}
   \toprule
    Property  & Present  & Previous   &  Previous \\
              & Simulations           &  Experiments       &  simulations\\
    \toprule            
    1--4 $\psi$\SI{}{\degree}  &\SI{100}{\degree},\SI{50}{\degree}\footnotemark[1]{}&\SI{115.1},\SI{48.4}{\degree}\footnotemark[2]{} & \SI{90}{\degree}, \SI{50}{\degree}\footnotemark[3]{} \\
    1--4 $\phi$\SI{}{\degree} &\SI{90}{\degree},\SI{100}{\degree}\footnotemark[1]{}  & \SI{103.6}{\degree},\SI{88.1}{\degree}\footnotemark[2]{} & \SI{90}{\degree},\SI{80}{\degree}\footnotemark[3]{} \\
                      
    1--6 $\psi$\SI{}{\degree}&\SI{90}{\degree},\SI{180}{\degree}\footnotemark[1]{}  &\SI{89.8}{\degree},\SI{180}{\degree}\footnotemark[4]{} & - \\
                      
    1--6 $\phi$\SI{}{\degree}&\SI{70}{\degree}\footnotemark[1]{}&\SI{70}{\degree}\footnotemark[4]{} & -\\
    
    $\rho$,\SI{}{\gram\per\cubic\centi\meter}&1.43\footnotemark[5]{}  &1.47\footnotemark[6] & 1.47\footnotemark[6] \\
                
    $T_\mathrm{g}$,\SI{}{\kelvin}&325\footnotemark[7]&300-340\footnotemark[8]{}&321\footnotemark[6] \\
                
    ${E_\mathrm{{TPS}}}$,\SI{}{\giga\pascal}&5.2\footnotemark[9]{}. &1.25\footnotemark[6]{}  &  4.3\footnotemark[6]{} \\
               
    ${E_\mathrm{{MMT}}}$,\SI{}{\giga\pascal}   & 171.95\footnotemark[9]{}  & & 171\footnotemark[10]{} \\
                
                 \bottomrule
 \end{tabular}  
 \renewcommand{\thempfootnote}{\arabic{mpfootnote}} 
\footnotetext[1]{The $\phi, \psi$ torsion angle distribution from the AA simulation of TPS melt at \SI{613}{\kelvin} and \SI{1}{\bar}}
\footnotetext[2]{XRD spectroscopy of hydrated cycloamylose \cite{Gessler1999}}
\footnotetext[3]{AA simulation of 9-mer Amylose in water \cite{Koneru2019}}
\footnotetext[4]{Replica exchange simulation and NMR spectroscopy on $\alpha$-1--6 glucopyranoside dimer \cite{Patel2014}}
 \footnotetext[5]{TPS melt density ($\rho$) calculated at \SI{300}{\kelvin} and \SI{1}{\bar}}
 \footnotetext[6]{70:30 ($wt$\SI{}{\percent}) amylopectin-sorbitol melt \cite{Ozeren2020, Ozeren2020a}}
 \footnotetext[7]{Glass transition temperature ($T_\mathrm{g}$) at the experimental cooling rate was obtained from the combination of temperature-density variation plot and WLF equation}
 \footnotetext[8]{Native barley starch having \SI{29}{\percent} glycerol \cite{forssell1997phase}}
 \footnotetext[9]{The Young's moduli ($E_\mathrm{MMT}$, $E_\mathrm{TPS}$) was calculated from the stress-strain curve at \SI{300}{\kelvin}}
 \footnotetext[10]{Density functional theory (DFT) calculation on MMT \cite{zartman2010nanoscale}}
 \end{minipage}
 \end{table}

\begin{figure*}[!htb]
     \centering
     \includegraphics[width=14cm,height=6.35cm]{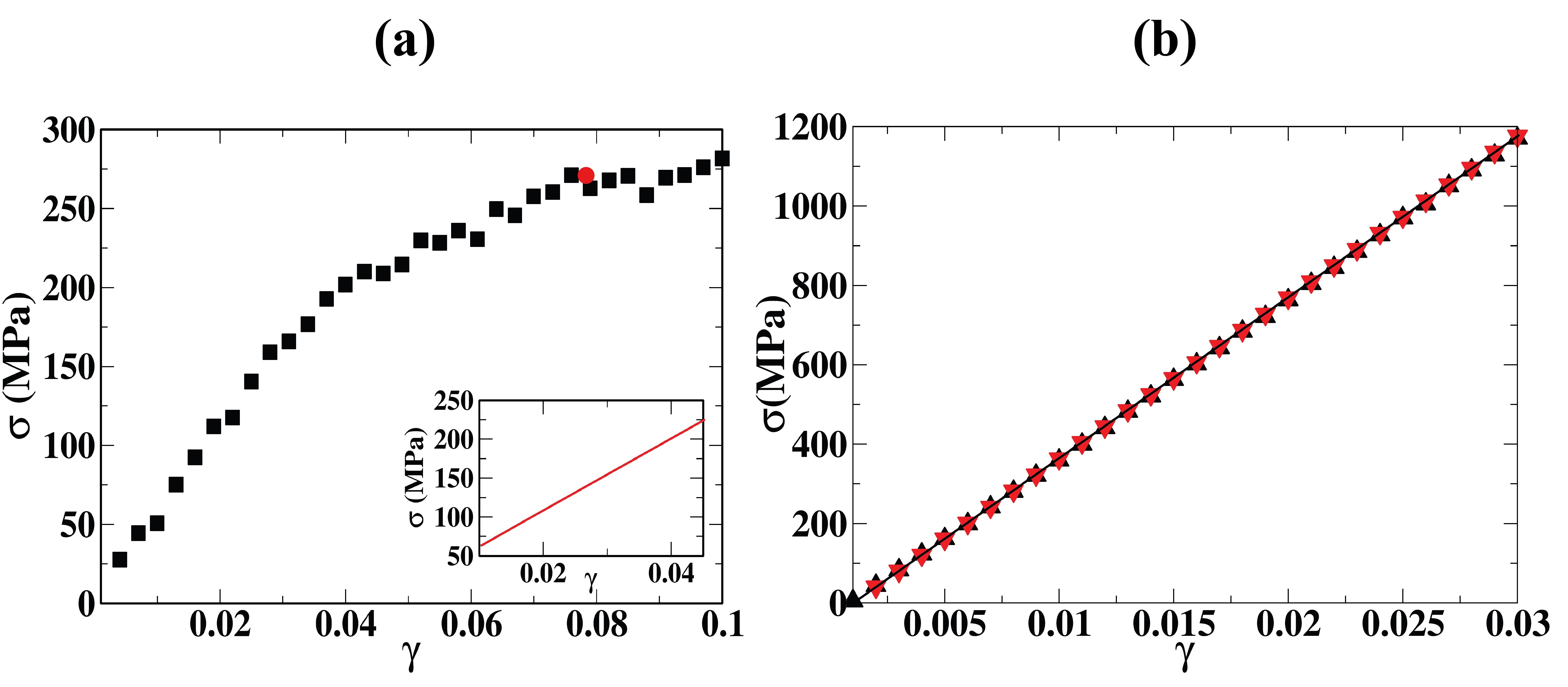}
     \caption{\textbf{Stress-strain behavior of the TPS melt and MMT sheet} \textbf{(a)} The linear part of TPS stress-strain data (represented by \redline) was used to estimate Young's modulus. The onset of the plateau (maximum stress) is represented by \redcircle. 
     (b) The AA (\blacktriiangle) and CG (\redinvtriangle) stress data of the TMA-MMT sheet. The system was stretched using a series of \SI{0.1}{\percent} unidirectional (in the $x$-direction) strains (at \SI{300}{\kelvin}, and \SI{1}{\bar}) and the corresponding Virial ($P_{xx}$) values at each step were used to estimate the stress.}
     \label{fig:stressstrain}
 \end{figure*}

\begin{figure*}[!htb]
     \centering
     \includegraphics[width=8.5cm,height=7.35cm]{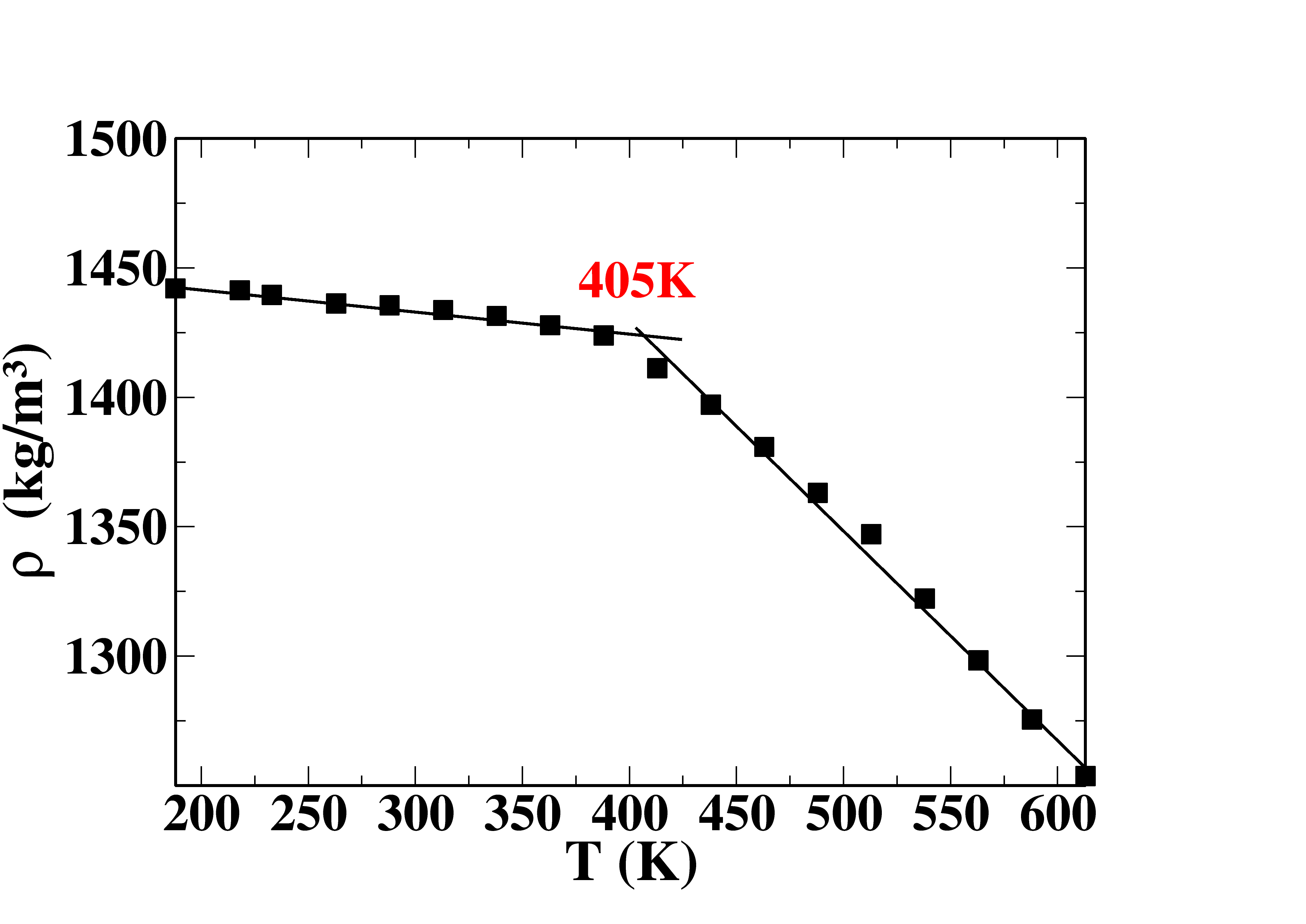}
     \caption{\textbf{Density-temperature plot for TPS melt} The TPS melt system was cooled from \SI{613}{\kelvin} to \SI{150}{\kelvin} in steps of \SI{25}{\kelvin}, where \SI{4}{\nano\second} NPT simulation was performed at each step, with last \SI{500}{\pico\second} trajectory at each step used for density calculation. The intersection point (highlighted in red) of rubbery and glassy region lines is used as an estimate of glass-transition temperature.} \label{fig:Tg}
 \end{figure*}

A TPS melt density of \SI{1.43}{\gram\per\cubic\centi\meter} at \SI{300}{\kelvin} compares well with a value of \SI{1.47}{\gram\per\cubic\centi\meter} obtained from both experimental measurements and simulations for a 70:30 ($wt$\SI{}{\percent}) amylopectin-sorbitol blend at \SI{300}{\kelvin} \cite{Ozeren2020, Ozeren2020a}. The available experimental data on the glass transition temperature ($T_\mathrm{g}$) and Young's modulus ($E_i$), both directly linked to the nature of intermolecular interactions and atomic-scale packing, allow an additional test for force field parameters and simulation convergence.  We obtained $T_\mathrm{g}=\SI{405}{\kelvin}$ from temperature-density variation obtained at a cooling rate of \SI{3.75e11}{\kelvin\per\minute} in AA simulation of TPS melt (Figure \ref{fig:Tg}), which yields a $T_\mathrm{g}=\SI{325}{\kelvin}$ on scaling to the experimental cooling rate of \SI{10}{\kelvin\per\minute} using the WLF equation (Equation \ref{eq:WLF}). This is in excellent agreement with a value of \SI{321}{\kelvin} obtained in a simulation of 70:30 amylopectin-sorbitol blend \cite{Ozeren2020a} and experimental estimates in range \SI{300}{\kelvin}--\SI{340}{\kelvin} for barley starch with \SI{29}{\percent} glycerol (a small molecule plasticizer like sorbitol) \cite{forssell1997phase}. $E_\mathrm{TPS}=\SI{5.2}{\giga\pascal}$ and $\sigma_\mathrm{max}=\SI{270}{\mega\pascal}$ obtained from the stress-strain variation for TPS melt at \SI{300}{\kelvin} (Figure \ref{fig:stressstrain}) compare well with $E=\SI{4.3}{\giga\pascal}$ and $\sigma_\mathrm{max}=\SI{285}{\mega\pascal}$ from AA simulation of 70:30 amylopectin-sorbitol blend at \SI{300}{\kelvin}, but is only within the same order of magnitude of the experimental estimate of $E=\SI{1.25}{\giga\pascal}$ \cite{Ozeren2020a}. The MMT sheet modulus of \SI{171.95}{\giga\pascal} from AA simulation matches the density functional theory (DFT) estimate of \SI{178.4}{\giga\pascal} \cite{zartman2010nanoscale} (both calculated at \SI{300}{\kelvin}). Table \ref{tab:AAproperties} gives an itemized comparison of various properties for TPS melt calculated in this work and those reported in other studies using experimental measurements and simulations. These favorable comparisons for the local conformation of the polymer chains as well as several macroscopic properties establish the high accuracy of TPS AA parameters used in this study.
\subsection{Development of CG Parameters}

The developed parameter set is intended for integration with the MARTINI-2 CG forcefield as available for several polymers and nanoparticles of interest, enabling direct application to their composites with TPS. We have not used the more recent MARTINI-3 CG parameterization \cite{souza2021martini}, since it does not yet have a polarizable model for water.\\

\subsubsection{CG Non-bonded Parameters for TPS melt}\label{melt non-bonded paramters}

The three diol groups of sorbitol (Figure \ref{fig:mapping} (c)) were modeled by a P4 bead with its self-interaction lowered from level I to level II ($\epsilon$=\SI{4.5}{\kilo\joule\per\mole}), since an unphysical aggregation was observed in a low-concentration aqueous solution \cite{sukenik2015osmolyte}. The refined polarizable model was used for water \cite{yesylevskyy2010polarizable, michalowsky2017refined}, where one neutral bead with self-interaction level V and cross-interaction level (\SI{60}{\percent} to \SI{95}{\percent} of standard MARTINI water P4 bead) is connected to two virtual beads ($q$=$\pm$0.46, no LJ interaction). This was shown to provide a better representation of dielectric behaviour of water. 
The initial bead choice in polymer (amylose and amylopectin) was taken from an existing set of MARTINI-2 CG parameters for polysaccharides (referred to as CGLITa), obtained by parameterization on AA simulations of mono- and disaccharides in polar and apolar solvents \cite{Lopez2009}. Accordingly, hemiacetal (Figure \ref{fig:mapping}(a): beads 3, 4, and 9), acetal (beads 1, 6, and 7), and diol groups (beads 2, 5, and 8) were modeled by P1, P2, and P4 beads, respectively. This is referred to as CGLITa parameter set. In this study, we used the melt density and the two-body excess entropy (higher values indicate stronger pair correlations) for polymer-polymer (given higher weight in evaluation) and polymer-sorbitol bead pairs to further optimize the bead-type assignments for amylose and amylopectin. 

The CGLITa melt density of \SI{1136}{\kilo\gram\per\cubic\meter} underestimates the melt density of \SI{1214.69}{\kilo\gram\per\cubic\meter} in AA simulations. The $S_2$ (Table \ref{tab:tbemelt}) for polymer-polymer beads is in excellent agreement with AA values (\SI{3}{\percent} higher), but there is a significant deviation for the polymer-sorbitol beads (\SI{108}{\percent} lower). An improved density estimate of \SI{1253}{\kilo\gram\per\cubic\meter} is obtained on using s-type for all polymer beads (CGLITa-S), a recommended choice for several ring molecules \cite{Marrink2007}. All $S_2$ values involving a polymer bead became higher (resulting from a weakening of interaction potential), leading to a poorer estimate for polymer-polymer beads (\SI{29}{\percent} higher than AA) but an improved estimate for polymer-sorbitol beads (\SI{11}{\percent} lower than AA). A similar improvement in density, but an incorrect prediction of the crystallization behavior of a highly concentrated sugar solution was observed on replacing regular beads with small s-type beads \cite{Lopez2009}. Alternatively, on using the s-type bead only for the hemiacetal group (referred to as CGLITb), as per the bead assignment used for lipopolysaccharides \cite{hsu2016molecular}, a density of \SI{1159.36}{\kilo\gram\per\cubic\meter} is obtained. This represented a small improvement over CGLITa, retained almost equivalent accuracy for polymer-polymer $S_2$, but polymer-sorbitol $S_2$ still showed a high deviation (\SI{86}{\percent} lower than AA). We made two additional changes in the bead assignments: acetal group as Na bead (recommended for obtaining accurate crystal behavior of cellulose \cite{wohlert2011coarse}) and one level reduction (from level I to level II) in the polymer-polymer diol bead (P4) self-interaction (shown to provide an accurate Virial coefficient and prevent unphysical aggregation of polysaccharides \cite{Schmalhorst2017,shivgan2020extending}). The parameter set obtained after these two changes in CGLITb is referred to as the CG1 parameter set (Table \ref{tab:CG1andCG2}). The CG1 melt density of \SI{1141.75} {\kilo\gram\per\cubic\meter} represented only a small improvement over CGLITa, a slightly increased polymer-polymer $S_2$ deviation (\SI{9}{\percent} lower than AA), and a somewhat improved polymer-sorbitol $S_2$ (\SI{61}{\percent} lower). Since CG1 provided the best balance in accuracy of $S_2$ for polymer-polymer and polymer-sorbitol pairs, we took it forward for further evaluation and optimization.

\subsubsection{CG Bonded Parameters for TPS melt}
\begin{figure*}[!htb]
     \centering
     \includegraphics[width=16.5cm,height=8.2cm]{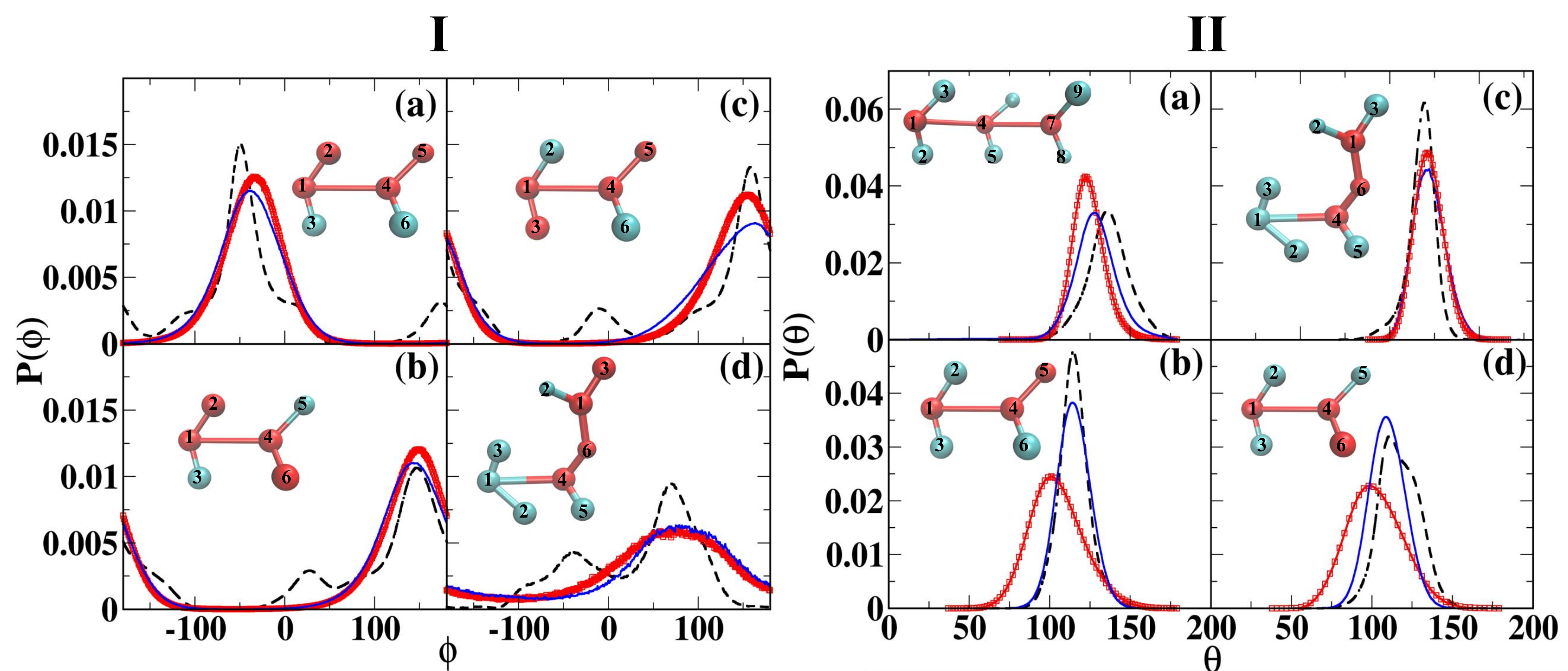}
     \caption{ \textbf{Bonded distribution for amylopectin} AA (\blackdotline), CG1 (\redwithsquare), and CG2 (\blueline) probability distributions, $P(\phi)$ for torsions ($\phi$), and $P(\theta)$ for three-bead angles ($\theta$) were estimated from TPS melt simulations at \SI{613}{\kelvin} and \SI{1}{\bar}. \textbf{I(a)} 2--1--4--5, \textbf{I(b)} 2--1--4--6, and \textbf{I(c)} 3--1--4--5 represent torsion distribution around the 1--4 linkage, and \textbf{I(d)} 3--1--6--4 around the 1--6 linkage. \textbf{II(a)} 1--4--7 and \textbf{II(b)} 4--6--1 represent angle distribution along 1--4 and 1--6 linkages, respectively. \textbf{II(c)} 1--4--5 and \textbf{II(d)} 1--4--6 represent intra-ring angle distributions. Bead numbers are as per Figure \ref{fig:mapping}. The beads involved in the calculated quantity are shown in red and rest in cyan.}
     \label{fig:d}
 \end{figure*}
\begin{figure*}[!htb]
     \centering
     \includegraphics[width=10cm,height=6.35cm]{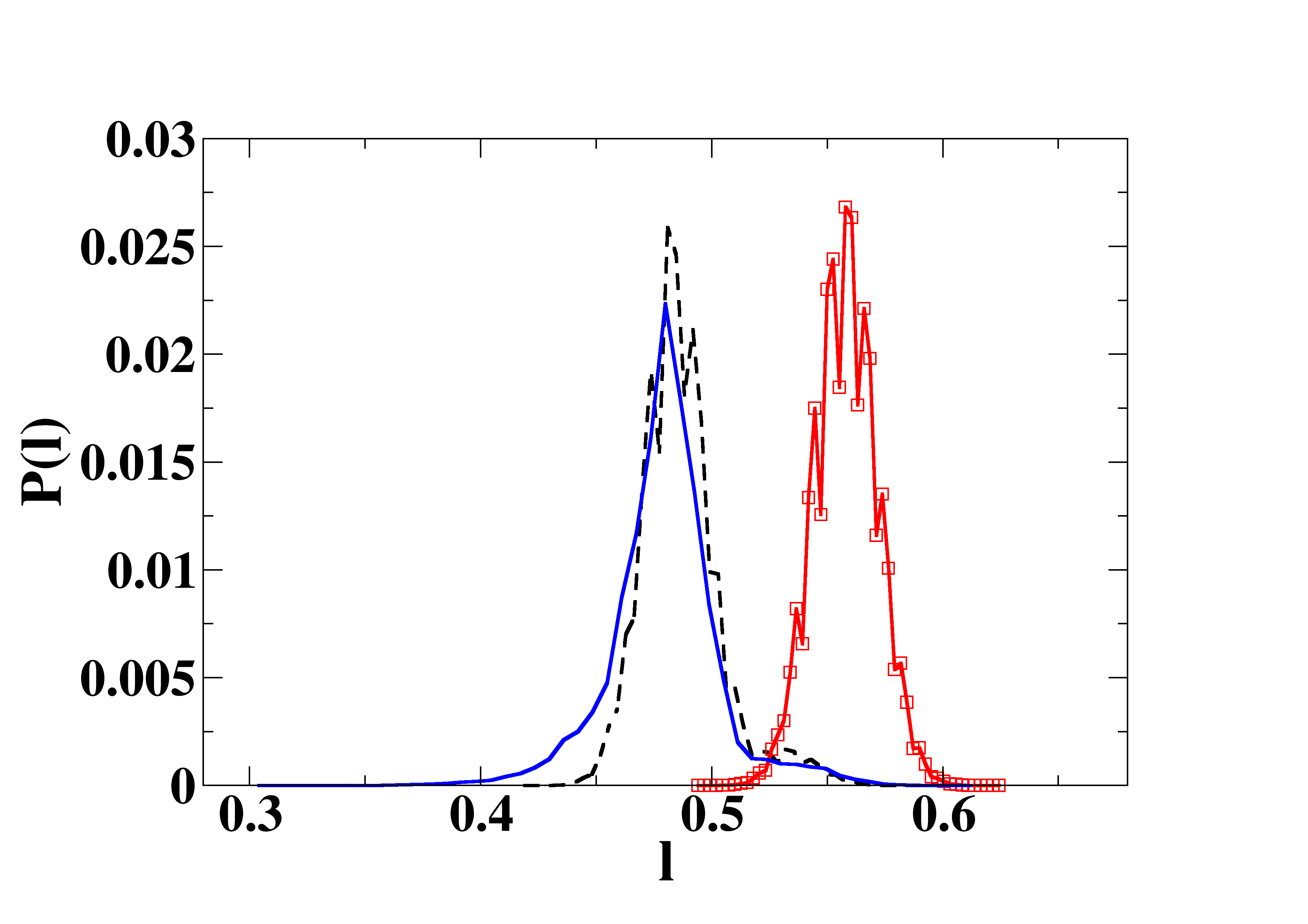}
     \caption{\textbf{AA and CG angle distribution of amylopectin}  y-axis show the probability distribution ($P(l)$) of 1--4 glycosidic bond ($l$) for AA(\blackdotline), CG1(\redwithsquare), and CG2(\blueline) models. The distribution obtained from TPS melt simulation at \SI{613}{\kelvin} and \SI{1}{bar}}.
     \label{fig:angdis2}
 \end{figure*}

The average 1–4 glycosidic bond length in AA simulation was \SI{0.48}{\nano\meter}, considerably smaller than \SI{0.56}{\nano\meter} in CG1, while the average 1-4-7 angle of \SI{125}{\degree} is close to the CG1 value for an amylose oligomer in an apolar solvent (\SI{120}{\degree}). Accordingly, we set $l_0=\SI{0.48}{\nano\meter}$ and $\theta_0=\SI{125}{\degree}$ for the corresponding bonded potential parameters in Equation \ref{eq:LJ}. Further, the force constant ($k_{\theta}$) for 1-4-5 and 1-4-6 angles needed to be doubled (w.r.t. to the CG1 value) to reproduce the AA angle distribution in TPS melt (Figure \ref{fig:d} (II)). This re-parameterized set, referred to as CG2 (Table \ref{tab:CG2bond}), provided an improved representation of the three-bead angle distributions for amylose and amylopectin in TPS melt (Figure \ref{fig:d} (II)) and \ref{fig:angdis2}). The torsion distribution for the 1-4 and 1-6 linkages are well-represented by a proper dihedral potential with multiplicity 1 (Figure \ref{fig:d} (I)) and are taken without change from CGLITa. Secondary peaks in the AA distribution, as seen here also, are not present in the CG distributions since the dihedral potential of multiplicity 1 is used in MARTINI FF to prevent overfitting. The CG2 parameter set also provided a slightly improved TPS melt density of \SI{1153.74}{\kilo\gram\per\cubic\meter} (comparable to CGLITb, better than CGLITa and CG1) and is the only set that provided a good estimate for $S_2$ of both polymer-polymer (\SI{2}{\percent} higher) and polymer-sorbitol (\SI{22}{\percent} lower). Further, amylose chain persistence length, $l_p$, of \SI{0.86}{\nano\meter} in AA simulations compares more favorably with CG2 (\SI{0.90}{\nano\meter}) than CG1 (\SI{0.96}{\nano\meter}). For comparison, simulations of amorphous amylose melt at \SI{713}{\kelvin} gave a value of \SI{0.9}{\nano\meter} with AA and \SI{1}{\nano\meter} with CG MARTINI \cite{gatsiou2022atomistic}. The chain radius of gyration, ${\langle R_\mathrm{g} \rangle}$, also follows the same trend for both amylose and amylopectin: within \SI{5}{\percent} of AA in CG2 and \SI{15}{\percent} of AA in CG1 (Table \ref{tab:conformation}). The 1--4 glycosidic bond length mismatch for the CG1 model (Figure \ref{fig:1416diah}(II)) is the likely reason for these small but consistent differences.
A lower angle stiffness and larger bond length for 1--4 bonds also lead to $\sim$ \SI{10}{\percent}-\SI{15}{\percent} higher $S_\mathrm{c}$ for amylose and amylopectin chains with CG1 parameters compared to AA (Table \ref{tab:conformation}). A correction for both bonded parameters in CG2 set gave $S_\mathrm{c}$ values within \SI{5}{\percent} of AA. Simultaneous improvements in several structural and thermodynamic properties make the CG2 parameter set a distinct improvement over existing parameters for TPS. 

\subsubsection{CG Non-bonded Parameters for TPS-MMT composite}\label{sec:CGnbDev}

\begin{table*}[!htb]
 \small
  \centering
  \captionof{table}{\textbf{Comparison of properties determined from AA and CG simulations of TPS melt and TPS-MMT-TMA composite.} The radius of gyration (${\langle R_\mathrm{g} \rangle}$), conformational entropy per CG bead ($S_\mathrm{c}$), and normalized diffusion coefficient of the polymer trimeric segments ($D_\mathrm{n}$) were calculated at \SI{613}{\kelvin} and \SI{1}{\bar}, with average and standard deviation calculated from \SI{900}{\nano\second} trajectories. The green, cyan, red, and blue values indicate a relative error (in percent) of $\leq$\SI{5}{\percent}, 5--\SI{10}{\percent}, 10--\SI{15}{\percent}, and $\geq$\SI{15}{\percent}, in CG values (w.r.t. AA), respectively.}
  \centering
  \label{tab:conformation}
\begin{threeparttable}
  
  \begin{tabular}{ c c c c c c c c c } 
    \toprule
    
      \multicolumn{2}{c}{} & \multicolumn{4}{c} {\textbf{Amylose}} & \multicolumn{3}{c}{\textbf{Amylopectin}}   \\
    \cmidrule(lr){3-5}\cmidrule(lr){7-9}
      \multicolumn{2}{c}{} & \multicolumn{1}{c}{\textbf{near}}  & \multicolumn{1}{c}{\textbf{far}}  &  \multicolumn{1}{c}{\textbf{melt}}  &   \multicolumn{1}{c}{} & \multicolumn{1}{c}{\textbf{near}} & \multicolumn{1}{c}{\textbf{far}} & \multicolumn{1}{c}{\textbf{melt}} \\
    \midrule
    ${\langle R_\mathrm{g} \rangle}$(\SI{}{\nano\meter}) &    AA & 1.35(0.09)    & 1.37(0.08)   & 1.38(0.02)  &  & 1.74(0.15) & 1.77(0.16) & 1.83(0.02)\\
            & CG1      & \textcolor{red}{1.49(0.18)}    & \textcolor{red}{1.57(0.10)}   & \textcolor{red}{1.58(0.05)}  &  & \textcolor{red}{1.95(0.25)} & \textcolor{red}{2.01(0.19)} & \textcolor{red}{2.07(0.03)} \\
            & CG1red      & \textcolor{red}{1.53(0.13)}    & \textcolor{red}{1.56(0.15)}   & \textcolor{red}{1.58(0.05)}  &  & \textcolor{red}{1.99(0.23)} & \textcolor{red}{1.99(0.23)} & \textcolor{red}{2.07(0.03)} \\
            & CG2   & \textcolor{green}{1.34 (0.19)}    & \textcolor{cyan}{1.45(0.12)}   & \textcolor{green}{1.43(0.03)}  &  & \textcolor{green}{1.78(0.19)} & \textcolor{green}{1.84(0.22)} & \textcolor{green}{1.83(0.06)} \\
            & CG2red    & \textcolor{green}{1.38(0.16)}    & \textcolor{green}{1.40(0.16)}   & \textcolor{green}{1.43(0.03)}  &  & \textcolor{green}{1.78(0.22)} & \textcolor{green}{1.84(0.21)} & \textcolor{green}{1.83(0.06)} \\
            \cmidrule(lr){3-9}
    ${\langle S_c \rangle}$  & AA  & 46.11(1.39)    & 46.73(1.64)   & 46.26(0.14)  &  & 45.82(2.70) & 46.63(3.10) & 46.23(0.11) \\
         $(\SI{}{\joule\per\mol\kelvin})$   & CG1       & \textcolor{green}{44.44(4.67)}    & \textcolor{green}{48.68(2.98)}   & \textcolor{red}{52.16(0.11)}  &  & \textcolor{green}{47.09(3.12)} & \textcolor{red}{51.41(3.54)} & \textcolor{red}{51.90(0.09)} \\
            & CG1red       & \textcolor{green}{48.23(1.69)}    & \textcolor{green}{48.31(2.52)}   & \textcolor{red}{52.16(0.11)}  &  & \textcolor{cyan}{50.30(3.61)} & \textcolor{red}{51.58(2.99)} & \textcolor{red}{51.90(0.09)} \\
            & CG2  &   \textcolor{cyan}{41.58(2.92)}    & \textcolor{green}{45.71(2.51)}   & \textcolor{green}{47.00(0.29)}   & & \textcolor{cyan}{43.25(2.21)} & \textcolor{green}{48.07(2.12)} & \textcolor{green}{45.59(0.25)} \\
            & CG2red  &   \textcolor{green}{44.53(2.17)}    & \textcolor{green}{45.18(3.18)}   & \textcolor{green}{47.00(0.29)}   & & \textcolor{green}{46.21(2.77)} & \textcolor{green}{48.21(2.44)} & \textcolor{green}{45.59(0.250)} \\
               \cmidrule(lr){3-9}
               & &  near &  overall  & $D_\mathrm{melt}$  & & near & overall & $D_\mathrm{melt}$\\
               \cmidrule(lr){3-9}
                ${\langle D_n \rangle}$  & AA  & 0.96    & 0.83  & 0.063(0.003)  &  & 1.16 & 0.79 & 0.033(0.003) \\
             & CG1       & \textcolor{blue}{-}    & \textcolor{blue}{0.47}   & 0.033(0.001) &  & \textcolor{blue}{-} & \textcolor{blue}{0.54} & 0.020(0.001) \\
            & CG1red       & \textcolor{red}{0.85}    & \textcolor{cyan}{0.78}   & 0.033(0.001)  &  & \textcolor{green}{1.11} & \textcolor{blue}{0.98} & 0.020(0.001) \\
            & CG2  & -    & \textcolor{blue}{0.41}   & 0.017(0.001)  &  & - & \textcolor{blue}{0.41} & 0.012(0.001) \\
            & CG2red  & \textcolor{cyan}{0.88}    & \textcolor{cyan}{0.92}   & 0.017(0.0010)  &  & \textcolor{blue}{0.97} & \textcolor{green}{0.79} & 0.012(0.001) \\
                   \bottomrule
  \end{tabular}
  \end{threeparttable}
\end{table*}

Three polymer-surface interaction dependent properties, viz. polymer normalized diffusion coefficient, $D_\mathrm{n}$, radius of gyration, $\langle R_g \rangle$, and conformational entropy, $S_c$, were simultaneously optimized to parameterize the cross-interaction for TPS-MMT-TMA beads. A lower $\langle R_g \rangle$ and $S_c$ near the MMT sheet in CG simulations compared to the bulk (not observed in AA) suggest a disc-like flattened configuration near the surface to maximize polymer-sheet contacts 
(Table \ref{tab:conformation}), which also gave a much lower $D_n$ for polymer trimeric segments and sorbitol molecules in CG simulations (Figure \ref{fig:normalizedoverallmsd} and Table \ref{tab:conformation}). This change in polymer chain conformation also affects the RDF and $S_2$ for all self- and cross-pairs of polymer and sorbitol beads, as discussed later in section \ref{sec:RDF_Comp}. All these observations indicate that MARTINI-2 forcefield parameters overestimate solid-fluid interactions for the TPS-MMT system. A reduction in solid-fluid interaction parameter ($\epsilon_{ij}$) was shown to be essential for accurate prediction of adsorption enthalpy of small organic molecules (butane, benzene, hexadecane), order-disorder transitions of hexadecane, and preferential adsorption of long-chain alkanes on graphite surfaces \cite{gobbo2013martini}. Here, the effective dispersive component of $\epsilon_{ij}$ was re-scaled for all TPS-MMT pairs as per Equation \ref{eq:redfact}, maintaining consistency with the parameterization strategy used for the TMA-MMT sheet \cite{Khan2019}. 
\begin{equation}\label{eq:redfact}
{\epsilon_\mathrm{i,j}}={\gamma(\epsilon_\mathrm{N0,N0})+(\epsilon_\mathrm{i,j}-\epsilon_\mathrm{N0,N0})},
\end{equation}
where, $i$ and $j$ represent TPS and MMT beads, respectively, and $\gamma$ is the scaling factor. A previous study has shown that taking the N0 bead self-interaction, $\epsilon_\mathrm{N0,N0}$, as purely dispersive and scaling the dispersive component for all interaction pairs as per Equation \ref{eq:redfact} gave an accurate distribution of polar and dispersive contributions to the cleavage free energy of two TMA-MMT sheets \cite{Khan2019}. $D_\mathrm{n}$, $\langle R_g \rangle$, and $S_c$ were calculated for the TPS-MMT system at six $\gamma$ values in range 0.15-1 (Table \ref{tab:allproperties} and Figure \ref{fig:normalizedoverallmsd}). The CG parameters were evaluated by using the property value in the composite normalized by the corresponding value in the melt (for the self-diffusion coefficient) or the property value in the near-sheet region normalized by the value in the far region (for the radius of gyration and the conformational entropy). This separated the effect of differences between AA and CG properties in the bulk systems (melt, far region) from the differences in the composite system, allowing specific optimization of TPS and MMT cross-interactions. For the CG1 set, $\langle R_g \rangle$ and $S_c$ ratios at $\gamma=0.21$ and $D_\mathrm{n}$ at $\gamma=0.25$ were closest to those in AA simulations, with $\langle R_g \rangle$ and $S_c$ at $\gamma=0.25$ and  $D_\mathrm{n}$ at $\gamma=0.21$ also in very good agreement with AA. A re-scaling factor of $\gamma=0.21$ was found most suitable for self-interaction of MMT beads \cite{Khan2019}, and therefore, the same factor was chosen here for the TPS-MMT cross-interactions for the CG1 parameter set (referred to as CG1red). For the CG2 parameter set, all three properties are in excellent agreement with AA at $\gamma=0.15$ and represent a distinct improvement over all other choices, and therefore, selected as CG2 parameter set for the composite system (referred to as CG2red).
  
The selected parameter set that concurrently optimize structural, thermodynamic, and dynamic properties are expected to provide an accurate estimate for other properties not used in optimization and also have better state-point transferability \cite{moore2014derivation, hu2019developing, banerjee2018coarse}. For example, the addition of miscibility data in MARTINI-3 FF development \cite{souza2021martini} allowed for a significant improvement in some of the key limitations of MARTINI-2 FF concerning large self-interaction of some species (such as polysaccharides) in solution \cite{Schmalhorst2017,shivgan2020extending}, poor reproduction of liquid properties at interfaces \cite{gobbo2013martini}.

\subsection{TPS-MMT Composite Properties from the CG Model}  

\begin{figure*}[!htb]
     \centering
     \includegraphics[width=10cm,height=12cm]{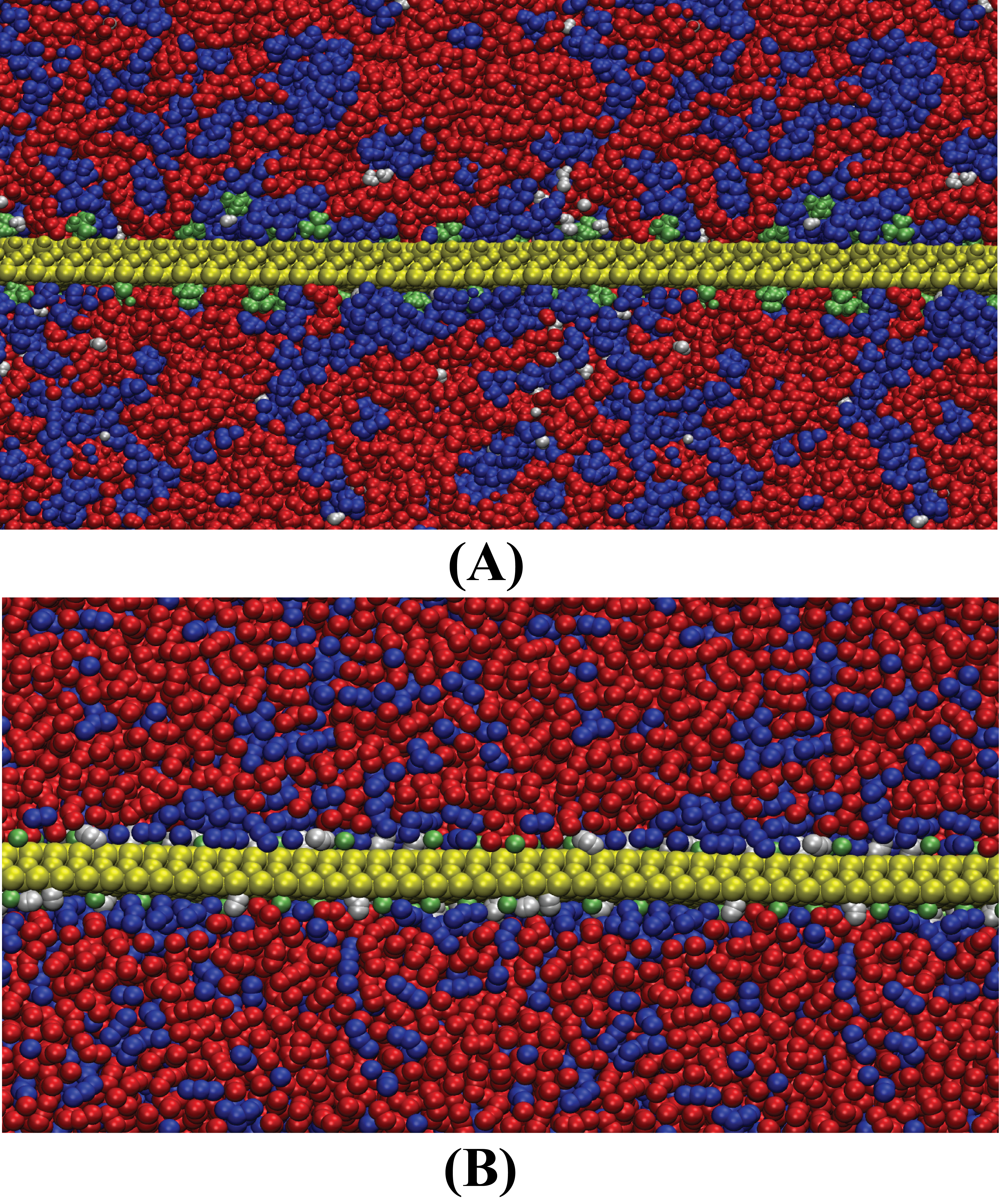}
     \caption{\textbf{AA and CG representation of TPS-TMA--MMT composite system}. \textbf{(a)} shows the AA representation of the composite system. \textbf{(b)} shows the same system in CG representation. The red, blue, yellow, and lime colors represent the polymer, sorbitol, water, and MMT, respectively.}
     \label{fig:AACGcomposite}
 \end{figure*}

Several properties of the composite system obtained using the different CG FFs were compared with the corresponding values obtained from AA simulations, and the inter-relationship between these properties was investigated. A single clay sheet sandwiched between two large TPS regions forms the model system \ref{fig:AACGcomposite}, which provides the simplest setup for separate determination of properties in the near-sheet and far-sheet ("bulk") regions while also allowing for establishing a chemical equilibrium between these regions.
\subsubsection{TPS Components' Density Profiles Above the TMA-MMT Sheet}

      \begin{figure}[!htb]
     \centering
     \includegraphics[width=8.25cm,height=7.0cm]{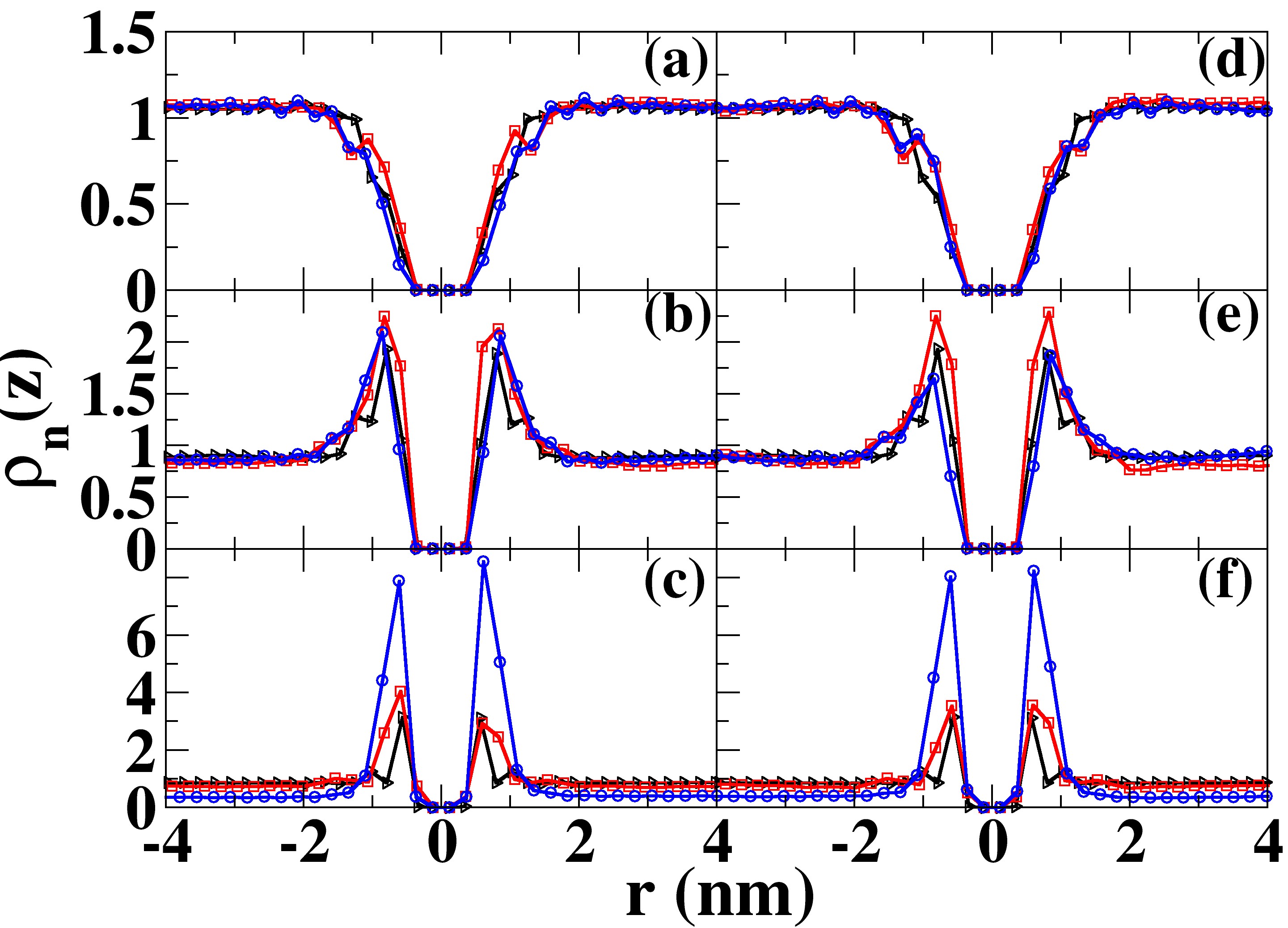}
    \caption{ \textbf{Density profile in TPS-MMT composite}. Normalized density perpendicular to the TMA-MMT sheet (placed in $xy$ plane), $\rho_\mathrm{n}(z)$, for \textbf{(a,d)} polymer, \textbf{(b,e)} sorbitol, and \textbf{(c,f)} water. CG parameter set 1 is in (a-c) and set 2 in (d-f), with AA as \blacktriianglen, CG as \redwithsquare, and CGred as \bluewithcircle.}
     \label{fig:de}
 \end{figure}
 
 The two plasticizers, water and sorbitol, show a sharp density peak near the TMA-MMT surface (Figure \ref{fig:de}), driven by the interaction between their highly polar hydroxyl groups and TMA-MMT silicate layer with mobile ions. This preferential adsorption of water and sorbitol leads to depletion of polymer chains at the sheet surface, but polymer density shows a sharp increase within the first solvation layer, increasing to $\sim \SI{50}{\percent}$ of the bulk value. This mixing of polymer and low molecular weight plasticizers in the interfacial region is possibly driven by the presence of polar beads, viz. diol (P4), hemiacetal (P1), and acetal (Na or P2), groups in the polymer chain. This provides one reason for an observed increase in TPS intercalation with increased water content in a TPS-cloisite (bentonite nanoclay) composite \cite{park2002preparation}. Figure \ref{fig:de}(a, d) shows that the scale of interactions (regular ($\gamma=1$) or reduced) have essentially no effect on the normalized density of polymer beads. This directly follows from high polymer depletion at the sheet surface, which reduces the direct effect of TPS-TMA-MMT bead interactions. The choice of CG model (CG1 versus CG2) also has a negligible effect, partly because of the density scaling by corresponding values in the melt system. Overall, all four CG models (CG1/2, regular/reduced) show an excellent agreement with CG and AA density profiles, $\rho_\mathrm{n}(z)$, for polymer and sorbitol. However, there is a significant increase in water density at the clay surface (Figure \ref{fig:de}(c, f)) for the CGred models. This implies that TPS-TMA-MMT cross-interactions are not fully optimized, with very low water content (\SI{1}{\percent}) possibly accentuating the effect of sub-optimal interactions. All component densities reach a plateau (close to the bulk value) at an average $z$-distance of \SI{2.5}{\nano\meter} from the TMA-MMT sheet, indicating minimal influence of the clay sheet beyond this distance. Therefore, the plane at $z=\SI{2.5}{\nano\meter}$ was used to distinguish between the near-clay and far regions for subsequent calculation of all reported properties.
\subsubsection{Polymer Chain Conformation and Dynamics}
Average chain conformation was characterized using radius of gyration, topological constraints and correlations using conformational entropy, and overall polymer dynamics using segment diffusion coefficients, all expected to show a significant dependence on polymer-surface interactions. In the AA simulations, the near-to-far ${\langle R_\mathrm{g} \rangle}$ and $S_\mathrm{c}$ ratios, and amylose and amylopectin trimeric segment $D_n$ were close to unity. It is expected that favorable polymer-surface interactions lead to chain expansion, lower $S_\mathrm{c}$, and lower $D_n$ in the near-surface regions (w.r.t. bulk) while unfavorable interactions lead to the opposite effect. For example, AA simulations showed that strongly attractive polymer-surface interactions for polyethylene oxide-Na-MMT \cite{suter2009computer} lead to doubling of ${\langle R_\mathrm{g} \rangle}$ and a decrease in diffusion coefficient in this system as well as in polyethylene-graphite composite \cite{daoulas2005detailed}. Weakly attractive interactions for polyethylene, polypropylene, and polystyrene near the TMA-MMT sheet lead to a $\sim \SI{10}{\percent}$ decrease in both $S_\mathrm{c}$ and diffusion coefficients \cite{Khan2019}. Conversely, the addition of repulsive nanoparticles to a LJ polymer accelerated chain dynamics \cite{smith2002molecular, varnik2002reduction} but had a negligible effect on ${\langle R_\mathrm{g} \rangle}$ \cite{karatrantos2015polymer,smith2002molecular}. For the TPS-MMT system, the interplay of polymer-surface (weakly repulsive), plasticizer-surface (attractive), and polymer-plasticizer (weakly attractive) interactions lead to polymer depletion at the clay surface and a sharp increase in the density in the first solvation shell of plasticizers at the surface (Figure \ref{fig:de}). This tempering of direct polymer-clay interactions and polymer-plasticizer mixing in the near clay region gave comparable ${\langle R_\mathrm{g} \rangle}$, $S_\mathrm{c}$, and $D_n$ values in the near-clay and the far regions. In a similar ternary composite system of cellulose nanocrystals (CNC)-polyvinylpyrrolidone (PVP)-polycaprolactone (PCL), PVP ${\langle R_\mathrm{g} \rangle}$ increased from $\sim$ \SI{0.90}{\nano\meter} (unfavorable PVP-CNC interactions lowered ${\langle R_\mathrm{g} \rangle}$ w.r.t. bulk) to \SI{1.65}{\nano\meter} on addition of dichloromethane (DCM), a good solvent for PVP that pushed the PVP chains away from CNC \cite{voronova2021interactions}. The use of standard TPS-MMT cross-interactions in CG1 and CG2 gave smaller near to far ${\langle R_\mathrm{g} \rangle}$ and $S_\mathrm{c}$ ratios than AA: in the range 0.92-0.97 for ${\langle R_\mathrm{g} \rangle}$ and 0.9-0.92 for $S_\mathrm{c}$ in CG compared to $\sim$ 0.98 for both ratios in AA. The $D_\mathrm{n}$ in the composite was significantly lower ($\sim 0.4-0.5$) than that in AA ($\sim 0.8-0.9$) (Figure \ref{fig:normalizedoverallmsd}). As discussed in Section \ref{sec:CGnbDev}, the solid-fluid interactions are typically overestimated in MARTINI FF \cite{yesylevskyy2010polarizable, gobbo2013martini}, leading to chain flattening and lower $S_\mathrm{c}$ and $D_\mathrm{n}$. Therefore, TPS-MMT cross-interactions were rescaled as per Equation \ref{eq:redfact}, which has built-in asymmetry leading to a smaller reduction in the pair interaction strength involving more polar beads.  An excellent agreement with AA values was obtained for ${\langle R_\mathrm{g} \rangle}$, $S_\mathrm{c}$, and $D_n$ at $\gamma=0.21$ for CG1 (referred to as CG1red) and $\gamma=0.15$ for CG2 (referred to as CG2red). The high level of rescaling is directly linked to the high covalent coordination state of the polymer chain and the MMT beads \footnote{A different value of re-scale parameter $\gamma$ for cross-interactions involving non-ring sorbitol and water beads would have been more appropriate but potentially affected transferability and therefore, was not pursued}, shown to significantly lower the dispersive interactions in earlier studies \cite{Khan2019, heinz2005force}. This use of multiple properties in CG parameter optimization is expected to provide improved state-point transferability as shown in previous studies, for example, use of multiple state point optimization \cite{moore2014derivation} or addition of density correction \cite{hu2019developing} in IBI-derived potentials, use of polymer hydration thermodynamics and structural properties in a CG FF combining MARTINI FF with IBI \cite{banerjee2018coarse}.  
\subsubsection{Radial Distribution Function and Two-body Excess Entropy}
\label{sec:RDF_Comp}

   \begin{figure*}[!htb]
     \centering
     \includegraphics[width=16.5cm,height=7.8cm]{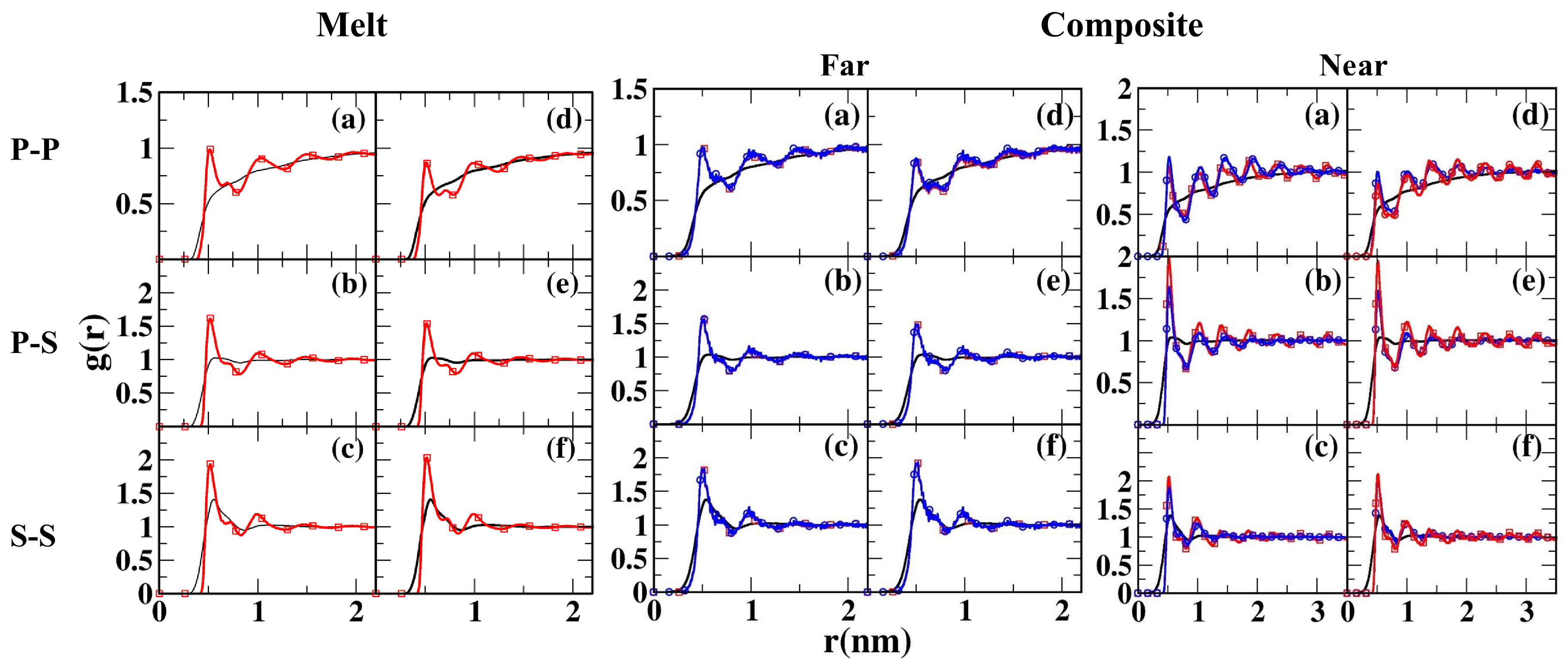}
     \caption{ \textbf{Radial distribution functions in TPS melt and TPS--MMT-TMA composite}. RDFs for polymer-polymer (P-P), polymer-sorbitol (P-S), and sorbitol-sorbitol (S-S) beads in the melt and in near- and far-regions in the composite (2D RDF): \textbf{(a-c)} $g_{\alpha\beta}$ for AA (\blackline), CG1 (\redwithsquare), and CG1red (\bluewithcircle), and \textbf{(d-f)} $g_{\alpha\beta}$ for AA (\blackline), CG2 (\redwithsquare), and CG2red (\bluewithcircle). Four bonded atoms were excluded from all RDF calculations.}
     \label{fig:RDF}
 \end{figure*}

  \begin{table}[!htb]
 \small
 \centering
  \caption{ \textbf{Normalized two-body excess entropy in TPS--TMA-MMT composite system.} ${S_{2,\alpha,\beta}}$ were calculated from the average RDFs shown in Figure \ref{fig:RDF}. The composite ${S_{2,\alpha,\beta}}$ values are reported after dividing with the respective ${S_{2,\alpha,\beta}}$ of TPS melt (Table \ref{tab:tbemelt}).  P and S refer to polymer (amylose and amylopectin) and sorbitol, respectively.}
  \label{tab:tbe}
      \begin{threeparttable}
  \begin{tabular}{c c c c c c c c}
    \toprule
      \multicolumn{2}{c}{Pairs} & \multicolumn{3}{c}{near} & \multicolumn{3}{c}{far}   \\
    \cmidrule(lr){3-5}\cmidrule(lr){6-8}
      \multicolumn{2}{c}{} & \multicolumn{1}{c}{P-P}  &  \multicolumn{1}{c}{P-S}  &  \multicolumn{1}{c}{S-S}  & \multicolumn{1}{c}{P-P}  &  \multicolumn{1}{c}{P-S} & \multicolumn{1}{c}{S-S}\\
    \midrule
    
    AA &           & 0.87    & 1.23  & 1.07 & 0.79 & 1.19 & 1.23  \\
    CG1 &      & 1.81    & 2.03  & 1.32 & 0.63 & 0.79 & 0.81   \\
    CG1red   &   & 1.69    & 1.05  & 0.83 & 0.60 & 0.79 & 0.81   \\
    CG2 &       & 2.00    & 2.60  & 1.16 & 0.60 & 0.91 & 0.86   \\
    CG2red    &       & 1.14    & 1.45 & 0.95 & 0.60 & 0.91 & 0.84   \\
    
    \bottomrule
  \end{tabular}
\end{threeparttable}
\end{table}

We have further evaluated CG1 and CG2 parameter sets by comparing estimates for radial distribution function and two-body excess entropy in the composite system to corresponding AA simulations, neither of which were used in parameter optimization. In the TPS-MMT composite system, the polymer-polymer (PP), polymer-sorbitol (PS), and sorbitol-sorbitol (SS) 2D $RDFs$ in the far region were identical to the melt (Figure \ref{fig:RDF}), indicating a negligible effect of clay on the structural properties of TPS at a distance (along sheet normal) of $\sim 2\langle R_\mathrm{g} \rangle$ from the sheet.  The corresponding RDFs for CG1 and CG2 parameters are in good agreement, with the size of the first solvation shell and the order of first-peak heights (SS > PS > PP) the same as AA simulations. However, at the full level of TPS-MMT interaction (used in CG1 and CG2), the near-sheet $RDFs$ show long-range structuring for all three pairs (Figure \ref{fig:RDF}). A 1:1 comparison of AA and CG RDFs is not easy since the peaks show a broadening in AA RDFs because of the use of COM of atoms making the CG bead. Therefore, individual pairwise ($S_\mathrm{2,\alpha,\beta}$) and total two-body excess entropy ($S_\mathrm{2}$), integrals over RDF calculated as per Equations \ref{eqn:tbe, eq:overalltbe, eqn:tbecomp}, were used for a quantitative comparison between AA and CG simulations (Table \ref{tab:tbemelt}). The near-region $S_\mathrm{2,\alpha\beta}$ values for CG1 and CG2 are up to \SI{108}{\percent} and \SI{130}{\percent} higher (in magnitude) than AA simulations, respectively.
An increase in peak height and long-range structuring were directly linked to highly attractive liquid-solid surface interactions \cite{smith2002molecular}, a known drawback of MARTINI 2 forcefield as discussed for CG parameterization. The use of CG1red and CG2red parameter sets lead to both a lowering of first-peak height and a reduction in long-range structuring in near-region RDFs (Figure \ref{fig:RDF}). 
 The relative errors (w.r.t AA) in near-region $S_\mathrm{2,\alpha,\beta}$ calculated for CG1red simulations were \SI{94}{\percent}, \SI{15}{\percent}, and \SI{22}{\percent} for polymer-polymer, polymer-sorbitol, and sorbitol-sorbitol respectively (an improvement of \SI{14}{\percent}, \SI{50}{\percent}, and \SI{1}{\percent} over CG1), while for CG2red simulations were \SI{31}{\percent}, \SI{18}{\percent}, and \SI{11}{\percent} (an improvement of \SI{99}{\percent}, \SI{93}{\percent}, and \SI{-3}{\percent} (this became slightly less accurate) over CG2), respectively. The success of the developed parameter set in accurately capturing diverse material properties beyond those used for optimization suggests their broad applicability for material characterization.
\section{Conclusions}
Enthalpy-based coarse-graining methods, such as DPD, require a state-dependent effective interaction parameter, wherein the phase space for its determination is significantly expanded for high system heterogeneity, for example, in the case of a TPS-MMT-TMA composite investigated in this work. Chemical-specific coarse-grained (CG) models, such as IBI, force-matching, or MARTINI, provide more effective parameterization for such systems. Here, we used AA simulation to predict several TPS properties, spanning from macroscopic (the glass transition temperature, Young's modulus) to microscopic (conformation along 1--4 and 1--6 glycosidic linkages). These properties were used as constraints in developing an accurate MARTINI-2 FF parameter set for TPS melt. While existing MARTINI FF parameters obtained using the data from AA simulations of mono- and disaccharides in polar and apolar solvents \cite{Lopez2009} form a good starting point, several important differences between AA and CG simulations were observed for chain conformation and thermodynamic properties of a TPS melt. We used the data from AA simulations in the present work and bead-type assignments for similar functional groups used in literature \cite{Lopez2009,hsu2016molecular,wohlert2011coarse,Schmalhorst2017,shivgan2020extending} to obtain a new parameter set for TPS melt, referred to as the CG2 parameter set. It included a shortening of the CG bond involving the 1--4 glycosidic link, an s-type bead for the hemiacetal group, Na bead type for the acetal group, one level reduction in the self-interaction of polymer diol bead, a small increase in the 1-4-7 angle from its value for amylose in apolar solvent, and a doubling of the force constants for 1-4-5 and 1-4-6 angles. The CG2 model provided simultaneous improvement in several structural and thermodynamic properties of TPS melt, making it a distinct improvement over existing MARTINI FF parameters for TPS.

The AA simulation of the TPS-MMT system revealed that the interplay of polymer-surface (weakly repulsive), plasticizer-surface (attractive), and polymer-plasticizer (weakly attractive) interactions lead to preferential adsorption of the plasticizer and depletion of polymer at the clay surface, and mixing of polymer and plasticizer within the first solvation shell, a feature captured very well by the optimized CG models used in this study.  This unique density profile also resulted in comparable $R_g$, $S_c$, and $D_n$ in near-clay and far regions, making it distinct from the expected behavior for attractive or repulsive polymer-surface interactions in simpler systems. However, the CG simulation of the TPS-MMT composite showed compaction and flattening of polymer chains (lower $R_g$) and significantly reduced dynamics (lower $S_c$ and $D_n$) in the near-sheet region. The use of liquid-liquid partition coefficients in MARTINI-2 parameterization makes it inherently less suitable for solid-liquid interfaces, with an unphysical freezing of liquid at a solid interface reported in several previous studies. In the present case, a high covalent coordination state for the polymer chains and MMT sheet, shown to significantly lower the dispersive interactions, further accentuated this effect. The downscaling of the effective dispersive component of the TPS-MMT MARTINI-2 interaction parameter led to simultaneous improvement in $R_g$, $S_c$, and $D_n$, with the selected CG2red parameter set giving $D_n$ for amylose within \SI{10}{\percent} of AA and all other values (for both amylose and amylopectin) within \SI{5}{\percent}. CG1red and CG2red models also significantly improved over the corresponding full interaction models (CG1 and CG2) in predicting two-body excess entropy (calculated from the radial distribution function). Thus, the developed MARTINI-2 CG parameters for TPS-MMT composites, which offer close to a hundred-fold speedup over AA simulations, can be used to accurately estimate properties not used in parameterization. 

We acknowledge that rescaling the sheet-TPS interaction is a sub-optimal solution that potentially limits transferability to systems with very different components. The optimal bead type assignments determined in this study, combined with a significant expansion of bead types and availability of scale factors for cross- and self-interaction in the recently released MARTINI-3 FF can allow for more rational development of parameters for various TPS-clay composites.

\begin{acknowledgement}
The authors thank the IIT Delhi High-Performance Computing (HPC) facility for providing the computational resources.
\end{acknowledgement}
\bibliography{achemso-demo}
\onecolumn

\section{Supplementary Material}
\doublespacing
\beginsupplement
\subsection{Temperature-pressure annealing cycles}\label{annealing}
\begin{figure*}[!htb]
     \centering
     \includegraphics[width=14cm,height=8cm]{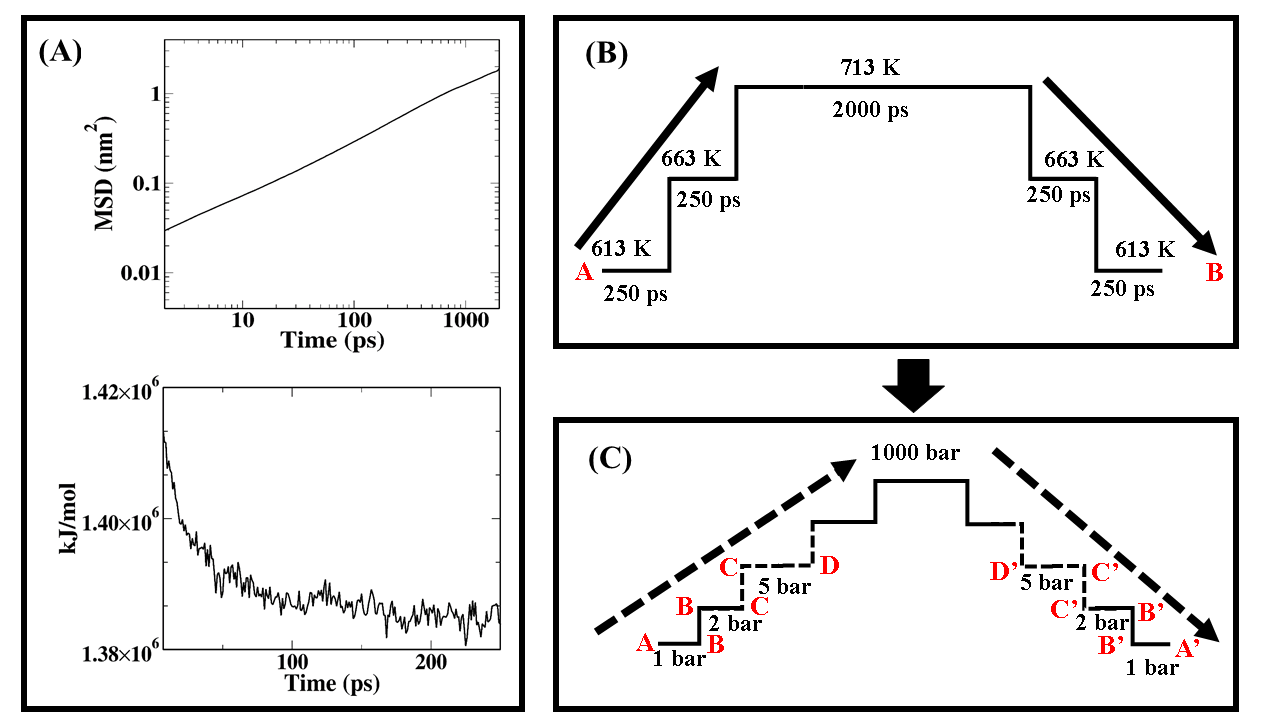}
     \caption{\textbf{Temperature-pressure annealing simulations} \textbf{A} The $MSD$ (at \SI{713}{\kelvin}) and potential energy plateau (at \SI{613}{\kelvin}), used to estimate simulation time and temperature in annealing cycle required in temperature annealing cycles (shown in (B)) to attain Brownian relaxation time and local equilibration of TPS melt, respectively. (\textbf{B}) In the temperature annealing cycle, states A and B represent the rescaled input and output simulation boxes, respectively. (\textbf{C}) T-P annealing cycle workflow along with the pressure values.} \label{fig:annealing}
 \end{figure*}

 The initial box, obtained using Packmol software, underwent a multi-step density increase followed by multiple temperature-pressure (T-P) cycles to obtain an equilibrated TPS structure free from the memory effect of the initial configuration.
 The coordinates of the simulation box were rescaled in steps of \SI{29.17}{\cubic\nano\meter} followed by energy minimization at each step using the steepest decent algorithm. The T-P annealing cycles were performed on the rescaled TPS system (state A) to accelerate the slow dynamics of the starch chains, leading to a more rapid attainment of the equilibrated configuration. (Figure \ref{fig:annealing}). 
 The simulation time at the lower (\SI{613}{\kelvin}) and higher temperature (\SI{613}{\kelvin}) of the annealing cycle was chosen to ensure local equilibration (potential energy plateau) and Brownian relaxation of the chain, respectively (Figure \ref{fig:annealing} (A)). Following the completion of each temperature cycle, the pressure was incrementally doubled. Once a pressure of \SI{1000}{\bar} was reached, the system was depressurized back to \SI{1}{\bar} using the same stepwise approach (Figure \ref{fig:annealing} (C)).
  The complete T-P cycles were repeated until the density of the initial (state A) and final (state A') states matched (Figure \ref{fig:annealing} (C)).

\subsection{Key Results}   
 
 \begin{figure*}[!htb]
     \centering
     \includegraphics[width=11cm,height=10cm]{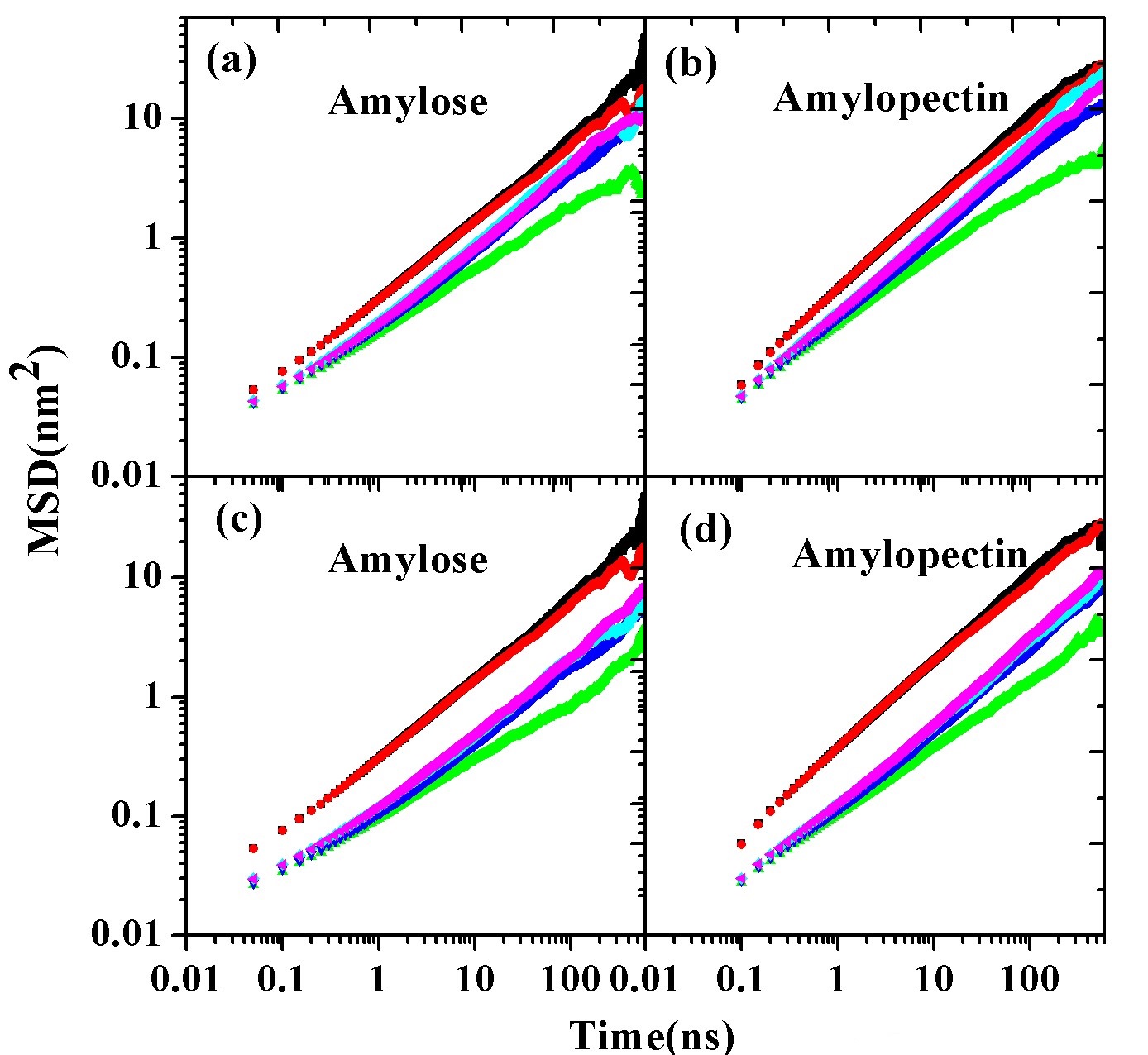}
     \caption{\textbf{MSD for TPS-MMT composite system } The MSD of amylose and amylopectin trimeric segments were estimated for CG1 (a,b) and CG2 (c,d) models. MSDs in Near-MMT regions for AA (\blacksequare), CG (\greentriiangle), and CGred (\megentasidetriiangle) and far-MMT regions for AA (\redcircle), CG (\blueinvtriangle), and CGred (\cyandiamond) are estimated from TPS--TMA-MMT simulation at \SI{613}{\kelvin} and \SI{1}{\bar}}.
     \label{fig:MSDnearfar}
 \end{figure*}
  \begin{figure*}[!htb]
     \centering
     \includegraphics[width=16cm,height=10cm]{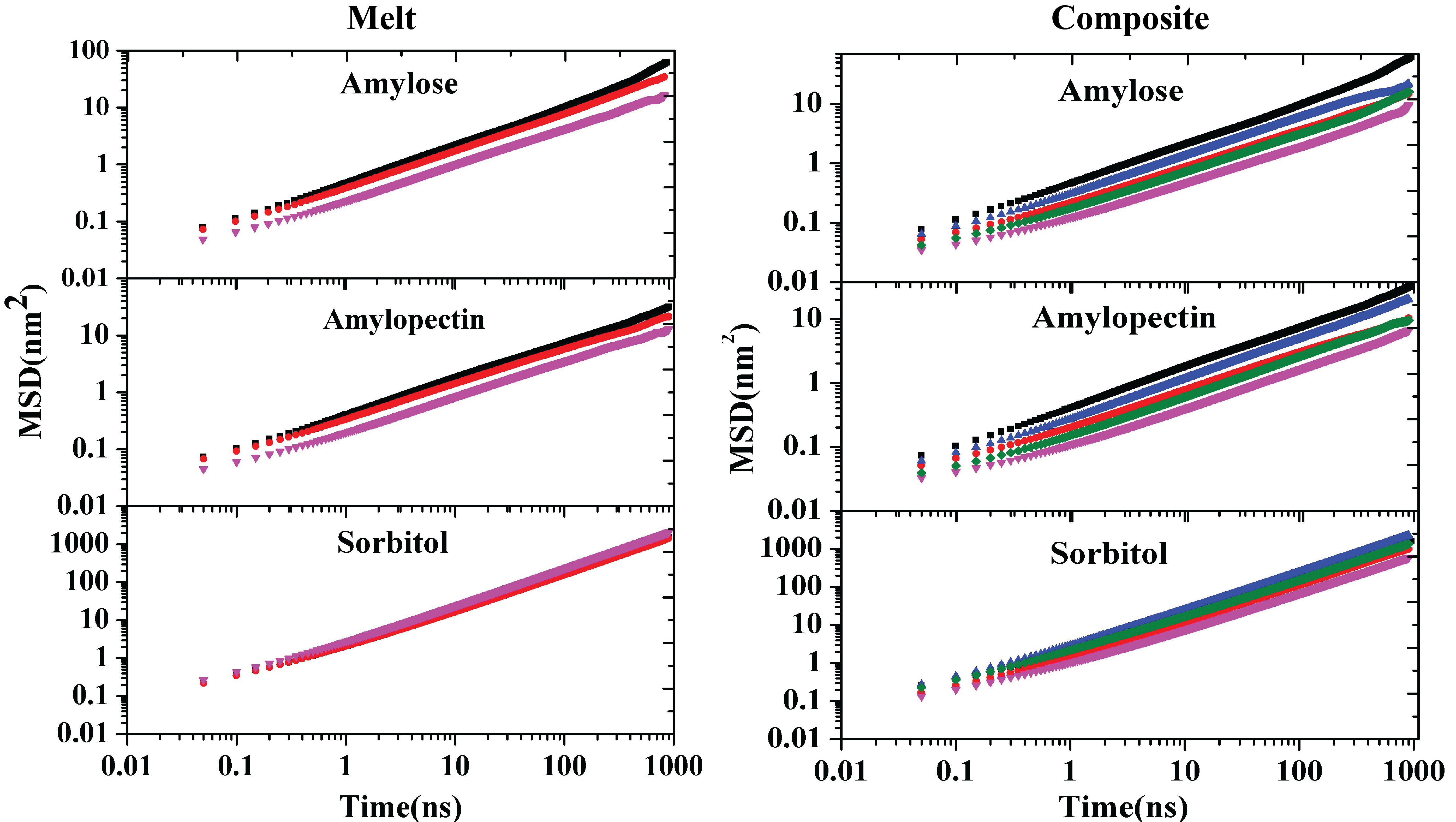}
     \caption{\textbf{MSD of TPS components in melt and composite systems} 
     MSD for AA (\blacksequare), CG1 (\redcircle), CG1red (\bluetriiangle), CG2 (\magentainvtriangle), and CG2red (\olivediamond) were estimated at \SI{613}{\kelvin} and \SI{1}{\bar}. In the composite system, the MSD was calculated in the $x$-$y$ plane using an equilibrated trajectory of \SI{900}{\nano\second} }.
     \label{fig:msdall}
 \end{figure*}

  \begin{table*}[!htb]
 \small
  \caption{ \textbf{Two-body excess entropy of TPS melt system} 
 The two-body excess entropies for polymer-sorbitol pairs were calculated by integrating the RDF (Figure \ref{fig:RDF}) of the corresponding pairs in the TPS melt, using Equation \ref{eqn:tbe}.}
  \label{tab:tbemelt}
      \begin{threeparttable}
  \begin{tabular}{c c c c c c c}
    \toprule
      \multicolumn{2}{c}{System} &&& \multicolumn{3}{c}{TPS melt}  \\
    \cmidrule(lr){1-7}
      \multicolumn{2}{c}{pairs} & &&\multicolumn{1}{c}{Polymer-Polymer}  &  \multicolumn{1}{c}{Polymer-Sorbitol}  &  \multicolumn{1}{c}{Sorbitol-Sorbitol}\\
    \bottomrule
    
    AA &         &&& -1.33    & -0.36  & -0.52\\
     CGLITa &      &&&  -1.29   & -0.75  & -0.57  \\
     CGLITb &      &&&  -1.39   & -0.67  & -0.61  \\
     CGSSa-S &      &&&  -0.95   & -0.40  & -0.95  \\
    CG1 &      &&&  -1.21   & -0.58  & -0.83  \\
    CG2 &       &&& -1.30    & -0.44  & -0.98  \\
    \hline
  \end{tabular}
\end{threeparttable}
\end{table*}
 The $S_{2,\alpha\beta}$ of polymer-sorbitol pair were calculated using Equation \ref{eqn:tbecomp}
\begin{equation}
   \mathrm{S_{2,\alpha\beta}}=-\pi\rho\int_{0}^{\infty}g_{\alpha\beta}(r)\mathrm{ln}\,g_{\alpha\beta} (r)-
   [g_{\alpha\beta}(r)-1]rdr
  \label{eqn:tbecomp}
\end{equation}

The diffusion coefficients were calculated using the Equation \ref{eq:MSD}. 
\begin{equation}\label{eq:MSD}
\mathrm{MSD (t)} = \mathrm{6 \cdot D \cdot t}
\end{equation}

 \begin{figure*}[!htb]
     \centering
     \includegraphics[width=8cm,height=10cm]{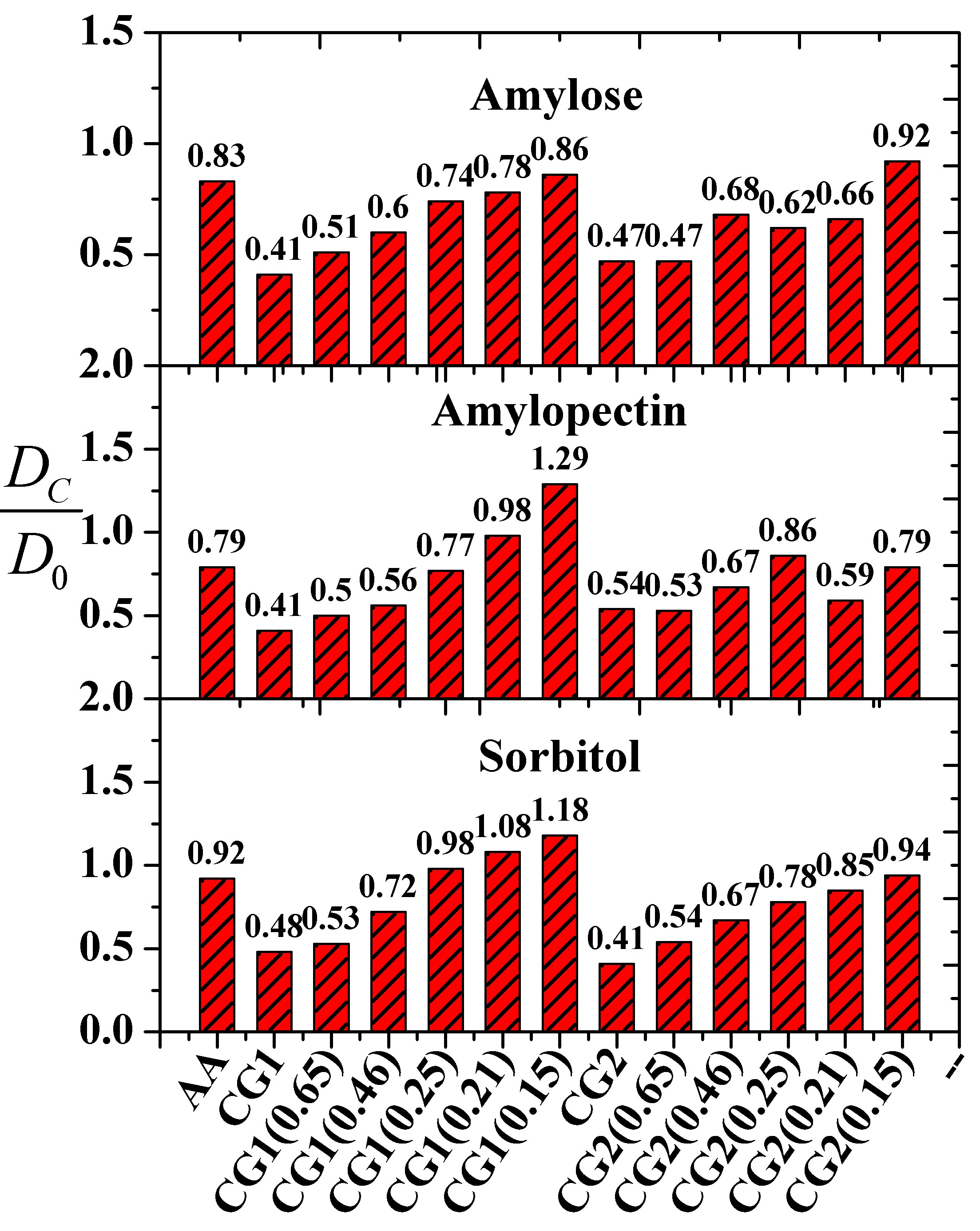}
     \caption{ \textbf{Normalized diffusion coefficient of TPS components} The diffusion coefficients of TPS components in melt ($D_0$) and composite ($D_c$) were obtained from the slope of ${MSD}$ (shown in Figure \ref{fig:msdall}) in the linear regime. The $MSDs$ were calculated from \SI{900}{\nano\second} long simulation trajectories, obtained at a temperature of \SI{613}{\kelvin} and a pressure of \SI{1}{\bar}}
     \label{fig:normalizedoverallmsd}
 \end{figure*}

 \subsection{CG-MARTINI parameters}
\begin{table*}[!htb]
  \centering
  \captionof{table}{\textbf{CG MARTINI bonded parameter 1-4 and 1-6 glycosidic links } The bonded parameter obtained from the AA simulation of TPS melt system at \SI{613}{\kelvin} and \SI{1}{\bar}. The columns show force constants $k_l$, $k_\theta$, and $k_\phi$ of bonds ($l_0$), angles ($\theta$), and dihedrals ($\phi_0$), respectively. Figure \ref{fig:mapping} illustrates the nomenclature used to represent beads.}.
  \centering
  \label{tab:CG2bond}
  \begin{threeparttable}

  \begin{tabular}{ c  c  c  c  c  c}
    \toprule
   
      \multicolumn{1}{l}{\textbf{type}} & \multicolumn{1}{l}{\textbf{topological pattern}}  &  \multicolumn{2}{c}{\textbf{parameters}} & \multicolumn{1}{c}{}& \multicolumn{1}{c}{}\\

    \midrule
               \textbf{bonds}   &   & \textbf{$l_o$}    &   \textbf{$k_l$ [\SI{}{\kilo\joule\per\mol\per\nano\meter\tothe{4}}]}&function& \\   \midrule
                      & 1-2   & 0.238    &   30000&1&\\
                      & 1-3   & 0.276    &   30000&1&\\
                     & 4-5   & 0.224    &   30000&1&\\
                      & 4-6   & 0.244    &   30000&1&\\
                     \textbf{1-4 link} & 1-4   & 0.482    &   30000&1&\\
                                  
                      \textbf{1-6 link}& 1-6   & 0.300    &   30000&1&\\
                      \midrule
                  \textbf{angle}   &   & \textbf{$\theta_0$}    &   \textbf{$k_\theta$ [\SI{}{\kilo\joule\per\mol}]}&function&\\
                  \midrule
                       & 2-1-4   & 70    &   100&2&\\
                       & 3-1-4   & 50    &   25&2&\\
                       & 1-4-5   & 140    &   100&2&\\
                       & 1-4-6   & 140    &   100&2&\\
                     \textbf{1-4 link}  & 1-4-7   & 125    &   180&2&\\
                       \textbf{1-6 link}& 4-6-1   & 100    &   55&2&\\
                       & 6-1-2   & 140    &   45&2&\\
                       & 6-1-3   & 140    &   35&2&\\
                       \midrule
                \textbf{dihedrals}   &   & \textbf{$\phi_0$}    &   \textbf{$k_\phi$ [\SI{}{\kilo\joule\per\mol}]}&function&multiplicity\\
                \midrule
                      & 2-1-4-5   & 120   &8 & 1&1\\
                       & 2-1-4-6   & -20   &5 &   1&1\\
                       & 3-1-4-5   & -40    &5&   1&1\\
                     \textbf{1-6 link}  & 4-6-1-2   & -100  & 5 & 1&1\\
                    \bottomrule
  \end{tabular}
\end{threeparttable}
 \end{table*}
The center of mass positions of functional groups (corresponding to CG beads) in AA trajectories were converted to pseudo-CG trajectories using Equation \ref{eq:mapping}. 
\begin{equation}\label{eq:mapping}
\mathrm{r^{CG}_{i}}=\frac{\sum_{j=1}^{n}{r_{j}}{m_{j}}}{\sum_{j=1}^{n}{m_{j}}}
\end{equation}
Where, $m_j$ and $\mathrm{r}_i$ are the masses and positions of atoms, $\mathrm{r^{CG}_{i}}$ are the positions for pseudo-CG beads.

 \begin{table*}[!htb]
 
  \centering
  \captionof{table}{\textbf{MARTINI non-bonded parameter of TPS melt}. The listed bead diameter ($\sigma_{ij}$) and interaction strength ($\epsilon_{ij}$) were used in stimulating the TPS melt system (for both the CG1 and CG2 models). The bead assignments are shown in Figure \ref{fig:mapping}. The P4c bead is a variant of P4 that employs a weaker level II self-interaction compared to level I in P4}
  \centering
  \label{tab:CG1andCG2}
  \begin{tabularx}{0.95\textwidth}{  X  X  X  X  }
   \toprule
      \multicolumn{2}{c}{Bead type} & \multicolumn{1}{l}{$\sigma$ (\SI{}{\nano\meter})}  &  \multicolumn{1}{l}{$\epsilon$ (\SI{}{\kilo\joule\per\mol})}  \\
    \multicolumn{2}{c}{} & \multicolumn{1}{c}{}  &  \multicolumn{1}{c}{}  \\
    \midrule
    
              P4c    & P4c  & 0.47    & 4.500 \\
              P4c    & Na  & 0.47   & 4.000\\
              P4c    & SP1  & 0.47   & 4.500 \\
              POL   & SP1  & 0.47   & 3.800 \\
              POL   & Na  & 0.47   & 3.325 \\
              POL   & P4c  & 0.47   & 4.275 \\
              POL   & POL  & 0.47   & 3.100 \\
              
                    \bottomrule
  \end{tabularx}

\end{table*}

 \begin{table*}[!htb]
 
  \centering
  \captionof{table}{\textbf{MARTINI non-bonded parameter of TPS-TMA--MMT composite system}. The Table illustrates the bead diameter ($\sigma_{ij}$) and the rescaled interaction strengths ($\epsilon_{ij}$) between TPS and MMT beads ($\gamma$=0.15). The SN0, SQ0, and SNa are MMT sheet beads representing $\mathrm{AlO_2H}$, $\mathrm{MgO_2H}$, and $\mathrm{SiO_2}$, respectively. }
  \centering
  \label{tab:sheetMARTINI}
  \begin{tabularx}{0.95\textwidth}{  X  X  X  X  }
   \toprule
      \multicolumn{2}{c}{Bead type} & \multicolumn{1}{l}{$\sigma$ (\SI{}{\nano\meter})}  &  \multicolumn{1}{l}{$\epsilon$ (\SI{}{\kilo\joule\per\mol})}  \\
    \multicolumn{2}{c}{} & \multicolumn{1}{c}{}  &  \multicolumn{1}{c}{}  \\           
    \midrule
                  SN0    & P1   & 0.47    & 1.025 \\
                  SN0    & P2  & 0.47    & 1.025 \\
                  SN0    & P4c  & 0.47    & 1.025 \\
                  SN0    & Na  & 0.47   & 0.525 \\
                  SN0    & SP1  & 0.43   & 0.769 \\
    
                  SNa    & P1   & 0.47    & 1.525 \\
                  SNa    & P2  & 0.47    & 1.525 \\
                  SNa    & P4c  & 0.47    & 1.525 \\
                  SNa    & Na  & 0.47   & 1.025 \\
                  SNa    & SP1  & 0.43   & 1.144 \\
    
                  SQ0    & P1   & 0.47    & 1.025 \\
                  SQ0    & P2  & 0.47    & 1.525 \\
                  SQ0    & P4c  & 0.47    & 2.625 \\
                  SQ0    & Na  & 0.47   & 1.025 \\
                  SQ0    & SP1  & 0.43   & 0.769 \\
                  SQ0    & SQ0  & 0.43   & 0.560 \\
                  SNa    & SNa  & 0.43   & 0.935 \\
                  SN0    & SN0  & 0.43   & 0.560 \\

                    \bottomrule
  \end{tabularx}

\end{table*}

\begin{table}[!htb]
 \small
  \centering
  \captionof{table}{\textbf{Comparison of AA and CG properties of TPS-MMT composite system} The radius of gyration (${\langle R_\mathrm{g} \rangle}$) and configuration entropy per CG bead ($S_\mathrm{c}$) were calculated for six different resealing factors ($\gamma$) in range of 0.15 to 1. All properties calculated at \SI{613}{\kelvin} and \SI{1}{\bar}.}
  \centering
  \label{tab:allproperties}
\begin{threeparttable}
  
  \begin{tabular}{ c c c c c c c c c c } 
    \toprule
    
      \multicolumn{3}{c}{} & \multicolumn{4}{c} {\textbf{Amylose}} & \multicolumn{3}{c}{\textbf{Amylopectin}}   \\
    \cmidrule(lr){3-5}\cmidrule(lr){7-9}
      \multicolumn{2}{c}{} &\multicolumn{1}{c}{\textbf{$\gamma$}} &\multicolumn{1}{c}{\textbf{near}}  & \multicolumn{1}{c}{\textbf{far}}  &  \multicolumn{1}{c}{\textbf{melt}}  &   \multicolumn{1}{c}{} & \multicolumn{1}{c}{\textbf{near}} & \multicolumn{1}{c}{\textbf{far}} & \multicolumn{1}{c}{\textbf{melt}} \\
    \midrule
    ${\langle R_\mathrm{g} \rangle}$(\SI{}{\nano\meter}) &    AA & & 1.35    & 1.37   & 1.38  &  & 1.74 & 1.77 & 1.83\\
    & CG1      & 1 &1.49    & 1.57   & 1.58 &  & 1.95 & 2.01 & 2.07 \\
    &       & 0.65 & 1.47	&1.58&	1.58 && 2.00	& 2.05&	2.07\\

    &      & 0.44 & 1.51 & 1.58&	1.58 && 2.01 & 2.03	&2.07\\
    &       & 0.25 & 1.52 & 1.58&	1.58 && 1.97&	2.01	& 2.07\\
    &       & 0.21 & 1.53	& 1.56&	1.58 && 1.99	&1.99	&2.07\\

    &       & 0.15 & 1.52	& 1.59&	1.58 && 2.03	&2.09	&2.07\\

    & CG2      & 1 & 1.34	& 1.45	& 1.43 && 1.78	&1.84	&1.83\\
    &       & 0.65 & 1.39	&1.47	&1.43 & &1.76 & 1.82 & 1.83 \\
    &       & 0.44 & 1.37	&1.43&	1.43& &1.77 &	1.83	& 1.83\\
    &       & 0.25 & 1.37&	1.41 &	1.43 & &1.79 &	1.85	& 1.83 \\
    &       & 0.21 & 1.40	&1.43&	1.43& &1.78 &	1.83	& 1.83\\
    &       & 0.15 & 1.38	&1.40	&1.43& & 1.78&	1.84 &	1.83\\

            \cmidrule(lr){3-10}
${\langle S_c \rangle}$  & AA  && 46.11  & 46.73 & 46.26  &  & 45.82 & 46.63 & 46.23 \\
$(\SI{}{\joule\per\mol\kelvin})$   & CG1 & 1     & 44.44    & 48.68   & 52.16  &  & 47.09 & 51.41 & 51.90 \\
&&0.65&45.95&	49.02&	52.16&& 47.07	&51.66	&51.90\\
&&0.44&44.81&	48.74&	52.16 && 48.71	& 51.93	& 51.90
\\
&&0.25&47.64	&48.65&	52.16&& 49.68&	51.61&	51.90
\\
&&0.21&48.23&	48.31&	52.16 && 50.30	& 51.58	& 51.90
\\
&&0.15&48.31&	48.54&	52.16 && 50.72&	51.99	& 51.90
\\

 & CG2  &  1 & 41.58    & 45.71   & 47.00   & & 43.25 & 48.07 & 45.59 \\
&&0.65&41.97&	46.23&	47.00 &&41.76&	48.45	&45.59\\
&&0.44&41.27&	47.03&	47.00 && 43.35	&49.23	&45.59\\
&&0.25&43.38&	45.96&	47.00&& 45.09&	48.12&	45.59\\
&&0.21&43.19&	45.96&	47.00&&45.79&	48.22&	45.59\\
&&0.15&44.53&	45.19&	47.00&& 46.21	&48.21	&45.59\\
 \bottomrule
  \end{tabular}
  \end{threeparttable}
\end{table}



\end{document}